\documentclass[12pt,pre,onecolumn,amsmath,amssymb,amsfonts,notitlepage]{revtex4-1}

\pdfoutput=1
\usepackage{dcolumn}
\usepackage{lipsum}
\usepackage{graphicx}
\usepackage{float}
\usepackage{bm}
\usepackage{color}
\usepackage{url}
\usepackage[bf]{subfigure}
\usepackage{rotating}
\usepackage{setspace}
\usepackage{hyperref}
\newcommand{\tbl}{\caption}
\definecolor{niceBlue}{RGB}{0,0,160}

\makeatletter
\renewcommand{\p@subsection}{}
\renewcommand{\p@subsubsection}{}
\makeatother

\makeatletter
\hypersetup{
		pdftitle={\@title}, 
		pdfauthor={\@author},
	    pdfcreator={Cesar Henrique Comin},
		colorlinks=true,       		
    	linkcolor=black,          	
    	citecolor=black,        		
    	filecolor=black,      		
		urlcolor=black,
		bookmarksdepth=4
}
\makeatother

\begin{document}


\title{Complex systems: features, similarity and connectivity}

\author{Cesar H. Comin$^1$}
\email[Corresponding author: ]{chcomin@gmail.com}
\author{Thomas K. DM. Peron$^1$}
\author{Filipi N. Silva$^1$}
\author{Diego R. Amancio$^2$}
\author{Francisco A. Rodrigues$^2$}
\author{Luciano da F. Costa$^1$}
\affiliation{$^1$Instituto de F\'{\i}sica de S\~{a}o Carlos, Universidade de S\~{a}o Paulo, S\~{a}o Carlos, S\~ao Paulo, Brazil\\
$^2$Instituto de Ci\^{e}ncias Matem\'{a}ticas e de Computa\c{c}\~{a}o, Universidade de S\~{a}o Paulo, S\~{a}o Carlos, S\~ao Paulo, Brazil}

\begin{abstract}
The increasing interest in complex networks research has been a consequence of several intrinsic features of this area, such as the generality of the approach to represent and model virtually any discrete system, and the incorporation of concepts and methods deriving from many areas, from statistical physics to sociology, which are often used in an independent way.  Yet, for this same reason, it would be desirable to integrate these various aspects into a more coherent and organic framework, which would imply in several benefits normally allowed by the systematization in science, including the identification of new types of problems and the cross-fertilization between fields. More specifically, the identification of the main areas to which the concepts frequently used in complex networks can be applied paves the way to adopting and applying a larger set of concepts and methods deriving from those respective areas.   Among the several areas that have been used in complex networks research, pattern recognition, optimization, linear algebra, and time series analysis seem to play a more basic and recurrent role.  In the present manuscript, we propose a systematic way to integrate the concepts from these diverse areas regarding complex networks research.  In order to do so, we start by grouping the multidisciplinary concepts into three main groups, namely features, similarity, and network connectivity.  Then we show that several of the analysis and modeling approaches to complex networks can be thought as a composition of maps between these three groups, with emphasis on nine main types of mappings, which are presented and illustrated.  For instance, we argue that many models used to generate networks can be understood as a mapping from features to similarity, and then to network connectivity concepts.  Such a systematization of principles and approaches also provides an opportunity to review some of the most closely related works in the literature, which is also developed in this article.
\end{abstract}

\maketitle

\linespread{1.0}

\section{Introduction}

The advances in computing along the last decades have strongly impacted the way in which science is done.  Not only much of the world has been mapped into data stored into databases and analyzed through statistics, but the very process of automatization has also implied in an ever increasing production of new information~\cite{Donovan:2008aa,Bell06032009}. At the same time that such advances have revealed the complex nature of our world, they also hold the promise for organizing and understanding this complexity. One aspect that has become clear by now is that it is not enough to study each concept or entity isolatedly in detail, characterizing the so-called reductionist approach.  As much important is the integration of such concepts and entities through relationships and connections, which is naturally provided by scientific areas focusing on connectivity, such as graph theory and complex networks -- it is hard to think of a discrete system that cannot be represented and analyzed in terms of connectivity and relationships.   The importance of such integration has been corroborated not only by an increasing number of related works, but especially by the variety of areas which are adopting these concepts and methods~\cite{costa2011analyzing,fortunato2010community}.

There is no single path to studying a system in terms of its connectivity. In some cases, one starts with the system and derives some of its characteristics, or features. In other circumstances, the focus is placed on the relationship between elements, such as while trying to predict how they originate and what the effect of their elimination would be.  Other studies concentrate on the time series produced by the individuals under analysis, while trying to identify joint variations.  Yet another approach is to devise models capable of producing specific features or behavior.  In spite of the seeming diversity of such approaches, there are elements which are common to most of them. At the same time, several of the concepts and methods adopted in complex networks are related or can benefit from toolsets of other areas, such as pattern recognition~\cite{duda2012pattern}, time series~\cite{hamilton1994time,makridakis2008forecasting}, statistics~\cite{feller2008introduction,reichl1980modern}, and visualization~\cite{borg2005modern}, among many others. For instance, the task of identifying clusters of objects in a given dataset, which is one of the main aspects of pattern recognition, can be related to community detection~\cite{fortunato2010community} in networks. Another example is the assortativity coefficient of networks~\cite{newman2002assortative}, which is based on the correlation coefficient commonly applied in time series analysis. The integration between such areas and approaches defines a potentially complex opportunity, involving a myriad of concepts and methods.

The identification of the shared elements between areas in an organized and systematic fashion would allow several benefits.  First, it would make clear what are the main methodologies involved.  Second, it would promote the cross-fertilization between methods which are shown to share several properties, in the sense that results and properties can be transferred from one to another.  In addition, the systematic identification of the basic elements could lead to new approaches for characterizing complex systems.

We consider that any proper representation of a complex system can be derived from the \emph{features}, \emph{similarity} and \emph{connectivity} of the elements contained in the system. The features representation concerns the characterization of the system components by a set of features $\mathcal{F}$, which define a feature space associated with the system. The choice of the relevant features to explain the system evolution is at the very core of creating a model of the system. The similarity representation involves portraying the system by the relationships between its elements, so as to allow the study of concepts such as the community structure and the centrality of the nodes. The connectivity representation deals with depicting the system by what are considered the relevant relationships between its elements. The three system representations are discussed in more detail in Section 2.

\begin{figure}[htbp]
\begin{center}
\includegraphics[width=12.5cm]{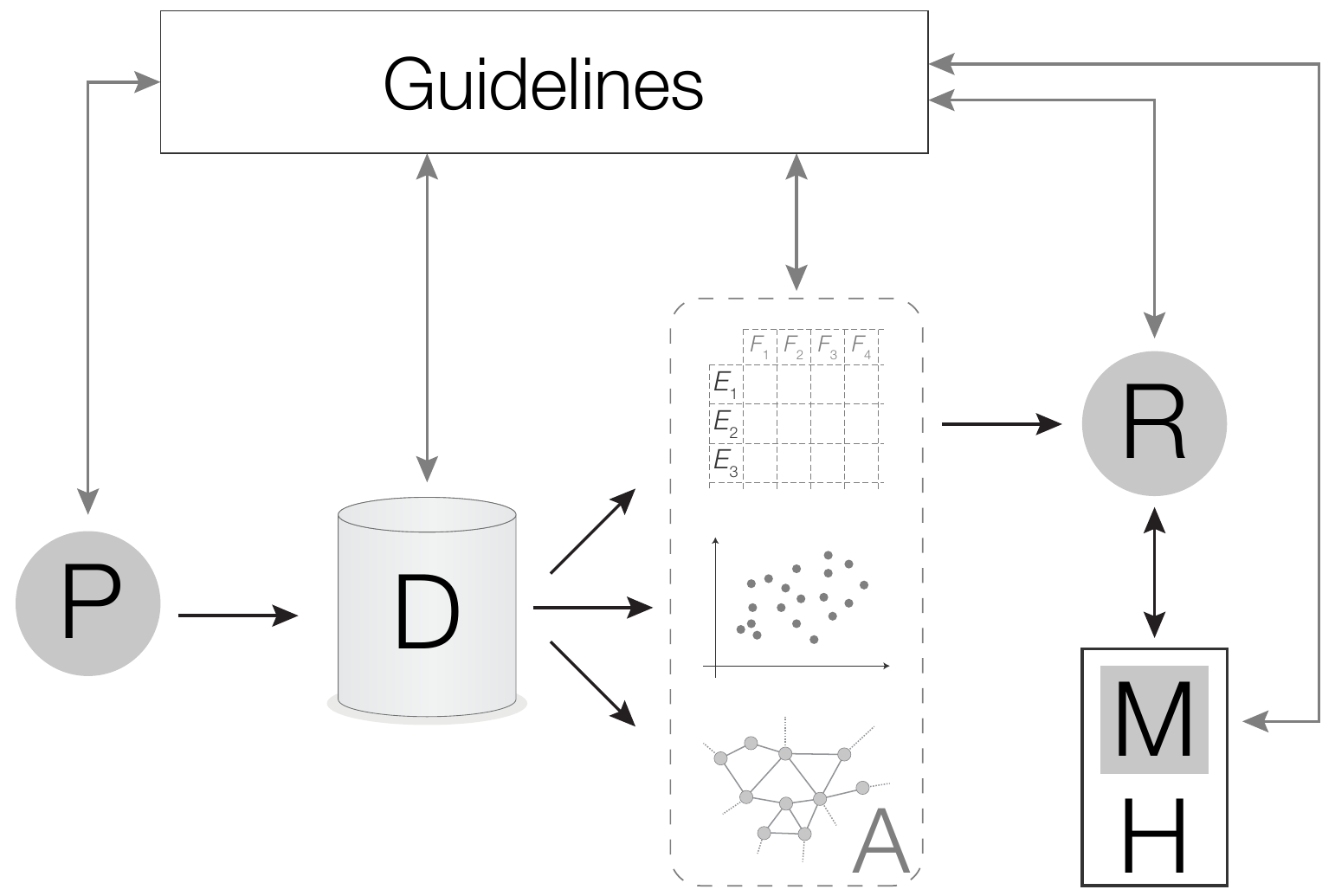}
\end{center}
\caption{Diagram illustrating the typical interaction between a problem (P), related database (D), alternative data representations (A), Researcher (R), methods (M), and implementation hardware (H). The choice of more specific configurations of these elements, especially the definition of the data representations, can be assisted by guidelines, subjected to intrinsic and extrinsic constraints and goals.}
\label{f:guidelines}
\end{figure}

The analysis of different representations of a system is an integral part of the scientific method, since the choice of the relevant variables and parameters to be investigated are a by-product of the considered representations. The systematization of the choices and methodologies involved in applying the scientific method characterize the field known as \emph{knowledge discovery in databases}~\cite{fayyad1996data} in computer science. This systematization implies a respective need for choosing and integrating the possible representations of a system, which is a challenging task given the broad scope of such a task. Important contributions to this concept were made using a set of methodologies comprising the so-called data mining~\cite{hand2001principles,han2001data} field. However, the main focus of the data mining approach to knowledge discovery has been on aspects such as multivariate statistics and data structure analysis, while lower effort has been put on considering the connectivity between elements in the system. Given the large growth of network theory over the past two decades, a new, connectivity-focused, approach to such a systematic application of the scientific method is needed. One of our objectives in the current work is to integrate such an approach. This is done by considering that a system can have the aforementioned three main representations, and methods aimed at better understanding the system correspond to mappings between these representations. The considered framework suggested and explored here is based on 8 guidelines, which are presented below. The  typical application of these guidelines is also illustrated in Figure~\ref{f:guidelines}.


\begin{enumerate}

\item \textbf{Problem demands:} In many situations, the problem may explicitly guide the choice of the methodologies and representations needed to analyze or model the considered system. For instance, the problem may require a scatterplot to visualize the relationship between two variables. Alternatively, the problem may need a table or list to compare the values of features among a few objects.

\item \textbf{Interactive exploration:} The interactive exploration of different representations of a system allows the researcher to choose one that is more suitable for solving the problem at hand. A common application of such a strategy is the use of interactive visualization software to find patterns in a dataset.


\item \textbf{Data filtering and selection:} Datasets may contain undesirable characteristics, such as redundancy, noise, and missing values. As a consequence, filtering or selection are procedures frequently employed in data analysis. For instance, by removing the redundancy of a dataset, one can achieve lower computational time and space to process and store a dataset. Such procedures can also be used to emphasize characteristics of interest in a dataset, for example, by removing noise from an input signal.

\item \textbf{Compatibility with researcher/field:} The expertise of a researcher or the tools commonly adopted in a research field usually require specific representations of a dataset. For example, in pattern recognition, one usually starts with tables describing the features of the objects being analyzed.

\item \textbf{Compatibility with methods:} Depending on the method being used to analyze the data, a particular representation may be required. For instance, if the method involves the calculation of shortest paths, a network representation is required.

\item \textbf{Compatibility with software/hardware:} The hardware or software involved in the given knowledge discovery process may also require proper representation of the data. For example, in array programming languages~\cite{shonkwiler2006introduction}, great optimization can be achieved by working with data organized as arrays.

\item \textbf{Complementary representations:} In the process of extracting
relevant information from data sets, different representations of the system under investigation can be explored so that further aspects of it are revealed.  For instance, provided a matrix comprising the geographical distances between
cities, one could analyze the spatial distribution of these elements. On the other hand, if the road network connecting the cities is given, questions regarding their connectivity and the distribution of shortest path lengths in such a network can be formulated.
 

\item \textbf{Cross-fertilization:} The search for proper data representations and methods that fit the above cited requirements can finally culminate in the cross-fertilization of techniques in different areas. For instance, pattern recognition methods can be used in the detection of modular structures in complex networks, which in turn can be employed in the context of machine learning problems. 

\end{enumerate}

\begin{figure}[htbp]
\begin{center}
\includegraphics[width=8.25cm]{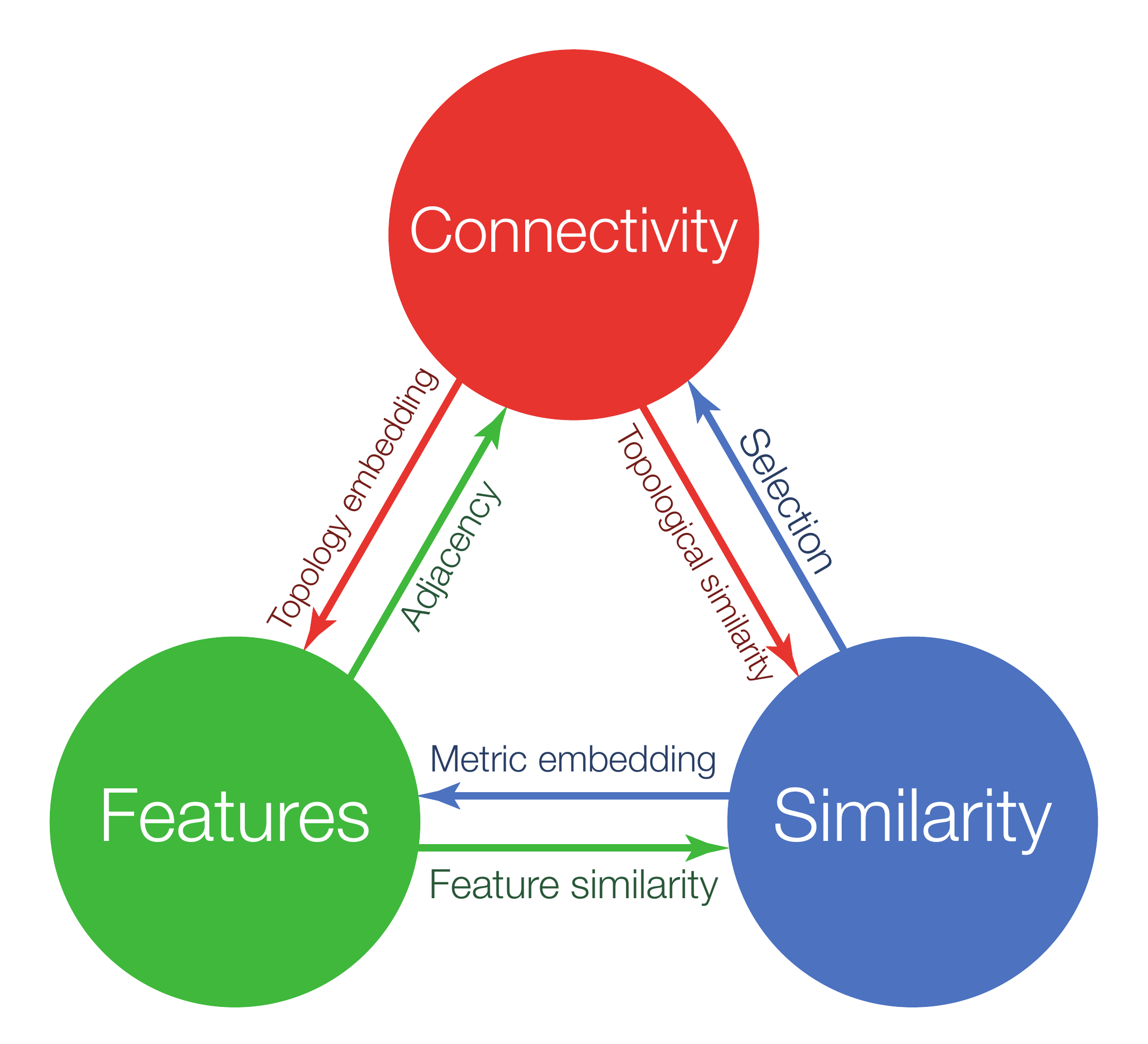}
\end{center}
\caption{Three main representations of a complex system, and their respective transformations.}
\label{f:triangle}
\end{figure}

\begin{figure}[htbp]
\begin{center}
\includegraphics[width=15cm]{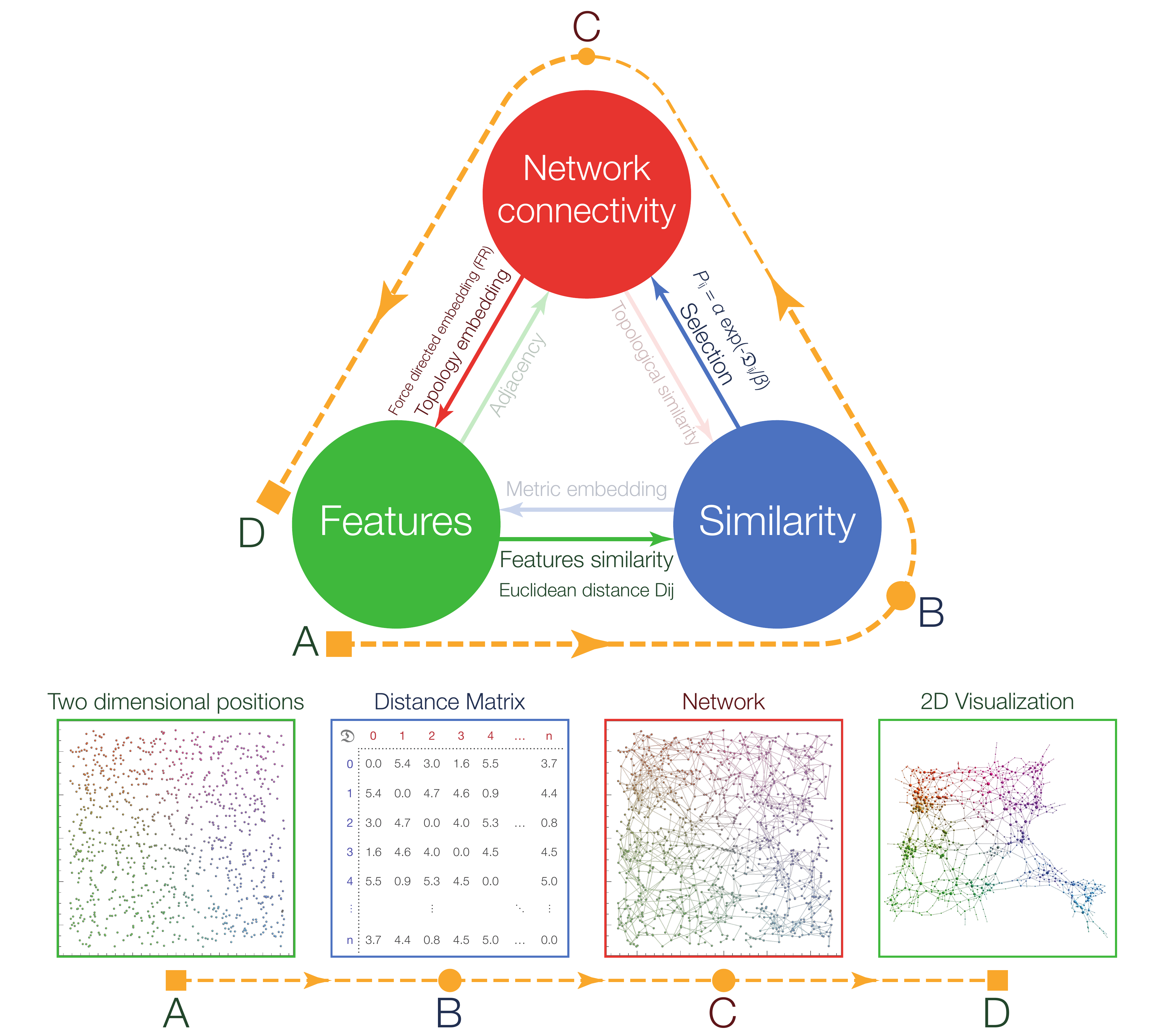}
\end{center}
\caption{Example of a transformation path from Figure~\ref{f:triangle}. Many network models follow this path. The case shown is known as the Waxman model~\protect\cite{waxman1988routing}.}
\label{f:triangle_wax}
\end{figure}

The framework considered in the current work is illustrated in Figure~\ref{f:triangle}. The feature, similarity and connectivity representations allow six immediate transformations. We note that such transformations can be \emph{complete}, in the sense that the mapping function is bijective, and therefore there is no information loss from the transformation, or they can be \emph{incomplete} (there is no inverse mapping). The current work concerns identifying and classifying a number of techniques described in the literature according to these six transformations, while also considering the aforementioned guidelines. Clearly, such techniques usually involve a combination of the six indicated transformations, that is, the system can undergo a \emph{path} along its three possible representations. As an example, we present in Figure~\ref{f:triangle_wax} the path followed by the Waxman network model~\cite{waxman1988routing}, which is commonly used in the study of spatial networks~\cite{barthelemy2011spatial}. In this model, the positions of a set of points are randomly drawn from a given range, which defines the features of the nodes, or equivalently, the feature space of the system, as shown in Figure~\ref{f:triangle_wax}A. Then, the Euclidean distance between each pair of nodes is taken, defining a distance matrix $D$ of the system, which is its similarity representation, as shown in Figure~\ref{f:triangle_wax}B. From the possible relationships between nodes, the Waxman model defines that we should select pairs of nodes $i$ and $j$ having large $P_{ij}=a\exp(-d_{ij}/\beta)$, where $P_{ij}$ is not a hard threshold, but a probability, thus obtaining the resulting network (Figure \ref{f:triangle_wax}C). We can think of an additional step to the Waxman transformation, which is visualizing the network by using a force-directed algorithm~\cite{Fruchterman1991Gr} in order to properly represent nodes that are topologically close to each other. This involves defining new features for the nodes, and the result is shown in Figure~\ref{f:triangle_wax}D.

\begin{figure}[!htbp]
\begin{center}
\includegraphics[width=0.9\linewidth]{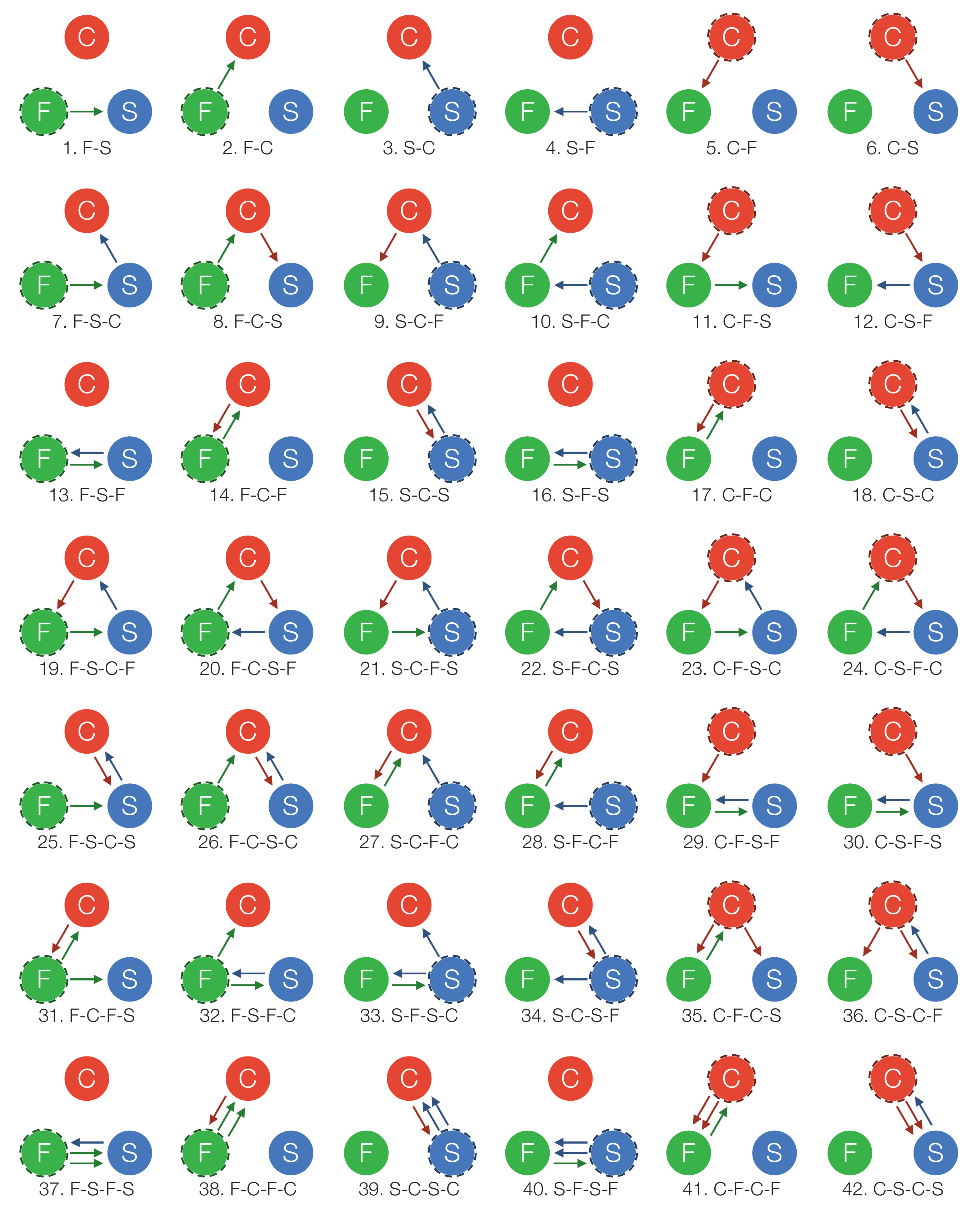}
\end{center}
\caption{Transformation paths among the three system representations depicted in Figure \ref{f:triangle}, when considering up to three transformations. Dashed outlines mean the initial representation of the system.}
\label{f:all_paths}
\end{figure}

For brevity, we henceforth represent the feature, similarity and connectivity representations by their respective first letters, F, S and C. A sequence of transformations, defining a path, is indicated by the respective sequence of their representations. For example, the path depicted in Figure~\ref{f:triangle_wax} is called a FSCF path. In Figure~\ref{f:all_paths} we show a catalog of what we consider being all possible paths that a system can undergo. More general cases, such as in a time-evolving network, can be seen as repetitions of a path shown in the catalog. The indicated paths are presented and associated with their respective methods throughout this work.

The paper is structured as follows. We first provide in Section 2 a more in-depth definition and discussion of the three main representations of a system. The six following sections, namely feature similarity, metric embedding, selection, topological similarity, topology embedding and adjacency are used to present and categorize many techniques described in the literature into the respective section. Although such techniques usually involve a number of transformations, most often one of the transformations can be regarded as being more relevant for the path, thus setting the proper section of the technique.

\section{Representations of a system}

In this section we briefly describe the three underlying domains of a complex system.

\subsection{Features}

A feature is any measurement used to characterize an object. Although they can be qualitative in nature, additional derivations require features to have a precise mathematical definition. Therefore, a set of quantitative features represent a mathematical description of an object. Here we focus on features that can be used to describe nodes in a graph. In this context, there are two fundamentally different types of features, intrinsic and induced. Features that are intrinsic to a node cannot be knowingly obtained from the connectivity of the network. For example, in a network representing social interactions, the age of a person could be, in principle, inferred from the connectivity pattern that this person makes, but this is hardly an attainable task. Since there is no known precise relationship between a topological measurement and the age of a person, the age is considered an intrinsic feature. Induced features are obtained in terms of topological properties of the node, some examples being the degree, the betweenness centrality and the clustering coefficient~\cite{costa2007characterization}. 



Features can have different scales. Intrinsic features are usually related to single nodes, while induced features can be used to characterize the immediate neighborhood of a node (e.g., degree and transitivity) or up to the entire network structure referenced at the node (e.g., betweenness, closeness and eccentricity). Another interesting concept related to features is the degree of completeness of the description that a set of features can provide about the node. When a set of features contains all the information about a node, being it intrinsic or induced, we call it a \emph{complete} set of features. Nevertheless, unless in some specific cases, the complete set of features usually contains an exceedingly large number of features, which makes working which such set unfeasible in practical cases. Therefore, the amount of features used in practice is always related to a balance between the level of description needed about the nodes and the maximum suitable number of features that can be handled during the analysis.

One special feature that we will extensively describe in this work is the spatial position of a node. If this feature is known for all nodes, the topology of the network can be analyzed as a function of spatial location or distances. Such relationship is usually influenced by other intrinsic or induced features, and the level of influence from other features on the position-topology relationship is a decisive factor of many characteristics of the network. One interesting example of such idea is the world-wide airport network~\cite{guimera2005worldwide}, where airports are considered as nodes and two nodes are connected if there is an airline route between them. In this network, the distance between airports bears influence on their connection probability. Furthermore, airports located in large cities have a higher chance of presenting long range connections~\cite{guimera2005worldwide,guimera2004modeling}.


\subsection{Similarity}
\label{s:similarity}
The similarity between two nodes in a network can be regarded as a scalar value indicating how close the two nodes are according to some criterion. Complex network theory usually deals with two main similarity classes, they are the \emph{features similarity} and \emph{correlation similarity}. The purpose of the features similarity is to associate a scalar to the relationship between values of a set of features. In many cases, this scalar is produced by using a dissimilarity measurement or distance in a feature space. In other words, a set of features characterizing the network nodes (e.g. age, degree, height) can be regarded as composing a metric space, and the nodes become points in this space. This process is usually called an \emph{embedding} of the nodes into a space. The most commonly used metric space is the Euclidean space~\cite{Berger1987}, mainly due to its intuitive relationship with the human perception. Yet, many other metric spaces can be used and a wide variety of real-world data and models are better embedded to certain non-Euclidean spaces, such as $n$-dimensional manifolds~\cite{tenenbaum2000global,roweis2000nonlinear,belkin2003laplacian}, elliptical~\cite{wilson2014spherical} or hyperbolic~\cite{bingham2000visualizing} spaces and even non-metric spaces~\cite{bronstein2006generalized}.

The other similarity class, here called correlation similarity, quantifies the level of dependence between variables associated to nodes. This dependence can be across time, over the feature space, or both. For example, we can represent companies by nodes and study the dependence between their stock values across a time interval (e.g., one month or year) or among other instantaneous features of a company, such as segment, market cap and the number of employees. The most widely used measurement of dependence is the Pearson correlation coefficient~\cite{snedegor1967statistical}, but many others exist \cite{smith_network_2011}. For example, the mutual information~\cite{cover2012elements} can be used to quantify non-linear dependencies between two variables.

Generally, similarity and dissimilarity are interchangeable through the use of simple transformations (Check Section~\ref{sec:SimilarityBasedModels} for more details). On the other hand, the transformation of a similarity or dissimilarity measurement into a distance, which defines a metric space, can represent a challenging task. This happens because metric distances must follow a strict set of formal mathematical rules~\cite{walter2012cl}. In the particular case of networks, a metric distance is a function $g:\mathcal{N}^2\rightarrow\mathbb{R}$ between nodes of a network $\mathcal{G}(\mathcal{N},\mathcal{E})$ constrained by the following properties:
\begin{enumerate}
\item {\bf axiom of coincidence, } $g(i,j)  \geq 0 \text{ if, and only if } i = j$;
\item {\bf the triangle inequality, } $g(i,j) + g(j,k) \geq g(i,k)$;
\item {\bf symmetry axiom, } $g(i,j)  = g(j,i)$;
\item {\bf non-negativity, } $g(i,j)  >  0 \text{ if } i \neq j$;
\end{enumerate}

Some of these constraints are usually relaxed to facilitate the process of finding a suitable definition of a generalized distance from a similarity measurement. This is the case of pseudometrics~\cite{howes2012modern}, in which the axiom of coincidence (1) is relaxed by allowing a null distance among pairs of distinct elements. Another common generalization is the use of semimetrics~\cite{wilson1931semi}, where the triangle inequality (2) property is not required. In essence, any deviation from the formal definition of a metric can lead to many distinct consequences, since it can severely affect the navigability and exploration in such spaces. For instance, this can undermine the performance of optimal path finder heuristics, such as the $A^*$ search~\cite{russell2009artificial}, which requires extra steps and memory space to account for distance functions not satisfying (2).

In most networks, two nodes are connected if they are similar or close in the aforementioned feature space. This similarity can be explicit (e.g., two airports are connected because they share an airline route) or hidden (e.g., an airport shutdown might cause an influx of planes to another airport, even though they do not share a direct route). In cases where the similarity is apparent, it can be used to construct a network directly from the nodes feature. Section~\ref{s:selection} explores some methods that can be applied for defining connectivity through similarity measurements.





\subsection{Connectivity}

In the previous sections we described how a given system composed by discrete parts can be embedded in a metric space through a set of intrinsic features. Hence, once the elements are completely mapped into this space, one can naturally define similarity measures, or conversely distances, quantifying how the parts relate to each other. This relationship pattern can then be used to define the \textit{network} representation of this system, characterized by a connection \textit{topology}. 

Before obtaining the network topology it is necessary to define certain criteria, given a set of similarity measures, by which the connections will be determined. In other words, the connections between pairs of nodes will be established according to some function that depends on the similarity (or distance) between them. How the selection of connections is done can dramatically change the network topology and will depend on the particular system that is being analysed and on the characteristics one is interested to study.  This is well exemplified by spatial random network models (described in Section~\ref{s:geometric_graphs}). Usually, one starts with $N$ disconnected nodes randomly distributed in a given metric space and considers that the probability of two nodes being connected depends on their geographic distance. As we shall see, choosing different spatial distributions and how the probabilities of connection depend on distances can generate structures ranging from Poisson random networks to scale-free ones. Furthermore, not only the selection criterion is crucial for the final network structure, but also the space in which the system is embedded (whether it is Euclidean or not, for instance).     

Naturally, one could also expect that this dependency of the connectivity pattern on the embedding space also holds for real-world spatial networks. The reason for that is simple: in real spatial networks every connection has a physical cost associated for its creation, which is intrinsically related to the system's geography~\cite{barthelemy2011spatial}. Examples of this interplay between connectivity and space can be found in, for instance, social networks~\cite{barthelemy2011spatial}, in which the probability of two individuals being connected decreases with the distance; power-grids~\cite{amaral2000classes,albert2004structural,sole2008robustness} and transportation networks such as roads and rail, all of them presenting strong geographical constraints~\cite{barthelemy2011spatial}.  Therefore, not only the space will impact on topological properties of these networks, but also on the performance of dynamical processes on them. 

\begin{figure}[!t]
\begin{center}
\subfigure[][]{\includegraphics[width=0.25\linewidth]{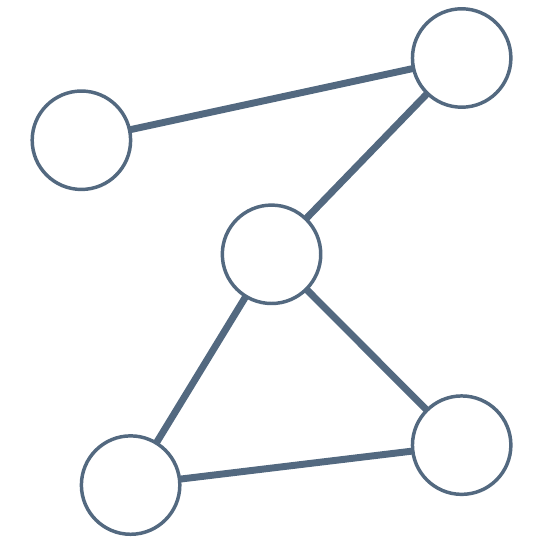}}
\subfigure[][]{\includegraphics[width=0.25\linewidth]{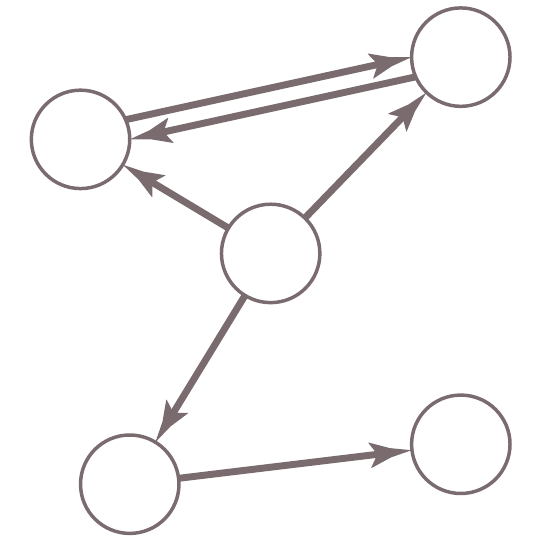}}
\subfigure[][]{\includegraphics[width=0.25\linewidth]{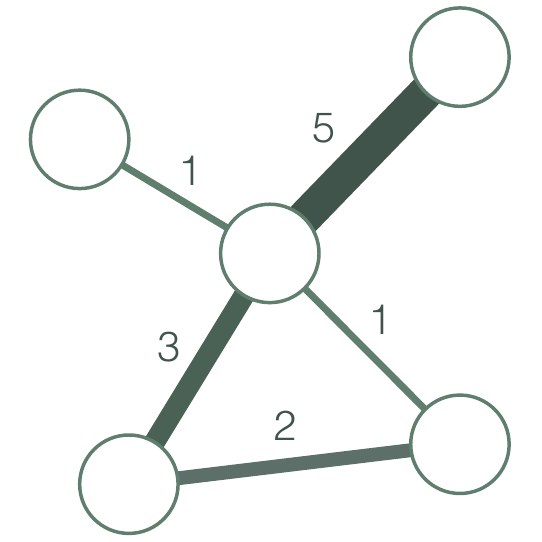}}
\end{center}
\caption{Examples of a (a) undirected (b) directed and (c) weighted network.}
\label{Fig:NetworkTypes}
\end{figure}

Formally, a network is represented by a graph $\mathcal{G}(\mathcal{N},\mathcal{E})$, where $\mathcal{N}$ is the set of nodes and $\mathcal{E}$ the set of edges. The mathematical entity that encodes network topology is the adjacency matrix $\mathbf{A}$, whose elements will represent the properties of the connections. More specifically, for \textit{undirected networks} $a_{ij}=a_{ji}=1$ if nodes $i$ and $j$ are connected ($(i,j)\in \mathcal{E}$) and $a_{ij}=0$, otherwise. For \textit{directed} networks, $a_{ij}=1$ if node $j$ has an incident edge 
departing from node $i$. Figs.~\ref{Fig:NetworkTypes}(a) and (b) depict examples of undirected and directed networks, respectively.  In these cases (Fig.~\ref{Fig:NetworkTypes}(a) and (b)), all connections are treated equally, however, in many applications, it is also relevant to assign intensity to the edges giving rise to weighted network. Examples of such cases can be found, for instance, in the Internet~\cite{pastor2007evolution} in which different levels of traffic between links can be observed; transportation networks~\cite{guimera2005worldwide,li2004statistical,barrat2004architecture} where the edges can quantify the flow of given quantities between two nodes; neural networks~\cite{latora2001efficient,bullmore2009complex} in which each edge has a different synaptic efficiency; and others~\cite{barrat2004architecture}. 
Similarly to unweighted cases, weighted networks can be represented by a graph $\mathcal{G}_{w}(\mathcal{N},\mathcal{E},\mathcal{W})$, where $\mathcal{N}$, $\mathcal{E}$ are the sets defined as before and $\mathcal{W}$ is the set of weights associated to the edges. Fig.~\ref{Fig:NetworkTypes}(c) illustrates a weighted network. 

\section{Feature similarity}

In some cases, the topology of networks can be used to express the similarity between nodes. In other words, two nodes are connected whenever they are similar with respect to the properties they share in the physical system
modeled by the network. For instance, if two individuals are friends in a social network, they possibly present a level of similarity with respect to certain social aspects. Now, this poses another question: given that we know intrinsic features of the nodes (not derived from the network topology), is it possible
to obtain the connectivity pattern from the quantification of the similarity of these features? The answer is that it may be possible, indeed. In this section we show how content similarity influences the
connectivity pattern in real networks by considering network models that are capable of describing structural properties of these systems. Subsequently, we discuss approaches to construct networks whose
topologies are inferred by quantifying statistical similarities between time series. Basically, all network models  and network approaches to time series described in this section will generally follow the FSC path 7 shown in Figure~\ref{f:all_paths}.



\subsection{Similarity-based models}
\label{sec:SimilarityBasedModels}

An interesting example of how content similarity influences networks formation is by Menczer~\cite{menczer2002growing}, where the author modeled the scale-free growth of the World Wide Web using lexical similarity between its pages. More specifically, the author proposed a generative network model in order to explain the WWW growth and topology based on the feature space in which the pages are embedded. Although some growth network models, such as the BA model, are capable of describing the power-law behavior of degree distributions of real networks~\cite{dorogovtsev2013evolution,newman2010networks}, they fail to describe the mechanisms of connections in the WWW. As Menczer remarked~\cite{menczer2002growing}, the BA model, for instance, has the bias that the older the node, the higher its degree. Moreover, the growth process requires global knowledge of the entire connectivity pattern, which turns out to be an unrealistic assumption to describe link formation in the WWW.

Therefore, in order to achieve a better description of link formation between Web pages, Menczer introduced a content-based generative model
by first defining the lexical distance between a given pair of pages  $(p_1,p_2)$~\cite{menczer2002growing}
\begin{equation}
r(p_1,p_2) = \frac{1}{s(p_1,p_2)} - 1,
\label{Eq:lexical_distance}
\end{equation}
where $s(p_1,p_2)$ is the cosine similarity given by
\begin{equation}
s(p_{1},p_{2})=\frac{\sum_{k\in p_{1}\cap p_{2}}w_{kp_{1}}w_{kp_{2}}}{\sqrt{\left(\sum_{k\in p_{1}}w_{kp_{1}}^{2}\right)\left(\sum_{k\in p_{2}}w_{kp_{2}}^{2}\right)}}
\label{Eq:cosine_similarity}
\end{equation}
and $w_{kp}$ is a given function for term $k$ in page $p$.

To model the preferential attachment in the WWW, Menczer defined the probability connection as a function of the lexical distance $r(p_1,p_2)$. More specifically, similarly to the BA growth processes, at each new step $t$ a new page $p_t$ is added and then $m$ new connections are made, linking $p_t$ to the $m$ previously placed nodes with probability given by~\cite{menczer2002growing}
\begin{equation}
\textrm{P}(p_{i},t)=\begin{cases}
\frac{k^{\textrm{in}}_i}{mt} & \textrm{ if }r(p_{i},p_{t})<\rho^{*}\\
c_{1}r^{-\alpha}(p_{i},p_{t}) & \textrm{ otherwise,}
\end{cases}
\label{Eq:Pr_menczer}
\end{equation}
where $k^{\textrm{in}}_i$ is the indegree of page $p_i$ at time $t$, $\rho^{*}$ the lexical distance threshold and $c_1$ and $\alpha$ constants. The choice for the probabilities in Equation~\ref{Eq:Pr_menczer} is motivated by the distribution of lexical distance~\cite{menczer2002growing}. Therefore, while the BA process assumes that each new node has global knowledge of network connectivity, the generative model described in Equation~\ref{Eq:Pr_menczer} requires only local knowledge about how the pages are connected. In other words, at time $t$, the probability of nodes $p_i$ and $p_t$ becoming connected will be proportional to the degree of node $p_i$ only if the distance between $p_i$ and $p_t$ in the similarity space is below $\rho^{*}$. This is a fair assumption, since it is reasonable to expect that the page's authors will intend to link their pages to others with similar content. The degree distribution of the simulated networks constructed through Equation~\ref{Eq:Pr_menczer} are in agreement with real data, showing very similar power-law exponents~\cite{menczer2002growing}.

The model presented in~\cite{menczer2002growing} was later improved by Menczer~\cite{menczer2004evolution}, where the author proposed the so-called \textit{degree-similarity mixture model}. Now, at time $t$, the $i$th page $p_i$ connects to the $t$th page $p_t$ with probability given by

\begin{equation}
\textrm{P}(p_i) = \alpha \frac{k^{\textrm{in}}_i}{mt} + (1 - \alpha){\textrm{P}_{\textrm{co}}}(p_i),
\label{Eq:Pr_menczer_II}
\end{equation}
where $\alpha \in [0,1]$ and
\begin{equation}
{\textrm{P}_{\textrm{co}}}(p_i)  \propto \left( \frac{1}{s(p_i,p_t)}  - 1\right)^{-\gamma},
\label{Eq:Pr_bar_menczer_II}
\end{equation}
where $s(p_i,p_t)$ is the cosine similarity between pages $p_i$ and $p_t$ and $\gamma$ is a constant~\cite{menczer2004evolution}. The new model was not just capable of reproducing the WWW's degree distribution
but also the similarity distribution as observed in the real network, complementing the approach presented in~\cite{menczer2002growing}.

The growth models defined in Equations~\ref{Eq:Pr_menczer},~\ref{Eq:Pr_menczer_II} and~\ref{Eq:Pr_bar_menczer_II} can 
be viewed as a generalization of the BA model specially devoted to describing the linking process in the WWW. The 
probability of receiving a new connection no longer depends solely on the node's popularity (degree) but also on the 
content similarity between the pages. The introduction of this dependence on similarity thus suggests that network 
models that establish some balance between popularity and similarity in node attractiveness have potential to better 
describe real-world networks. In fact, recently, Papadopoulos et al.~\cite{papadopoulos2012popularity} proposed a 
growing network model that  take precisely these properties into account, being able to properly describe the 
evolution of different real-world network with great accuracy. Besides describing the connection probability in these 
systems, other mechanisms of network formation such as preferential-attachment and fitness 
models~\cite{bianconi2001competition} (see also Section~\ref{sec:Fitness}) turn out to be particular cases of the more 
general approach presented in~\cite{papadopoulos2012popularity}. The relevance of content has also been reported 
in models of citation networks~\cite{Amancio2012427}. It has been shown that connectivity between papers is indeed 
based on content similarity, but different research areas tend to follow distinct connectivity rules for similar 
papers~\cite{Amancio2012427}. The surprising absence of citations between similar papers was quantified 
in~\cite{amancio2012using}.

In order to model a growth process having a competition between popularity and similarity to
attract new connections, to each new node $t$ $(t=1,2,...)$ it is assigned the polar coordinates $(r_t,\theta_t)$ in the
feature space. The term corresponding to the popularity of the nodes is the radial coordinate, which evolves in time according to $r_t = \ln t$~\cite{papadopoulos2012popularity}, and the angular coordinate $\theta_t$ is randomly drawn. The new node will connect to the $m$ closest nodes that minimize
the hyperbolic distance $x_{st} = \ln r_s + \ln r_t +  \ln(\theta_{st}/2)= \ln(s t\theta_{st}/2)$, where $(r_s, \theta_s)$
is the coordinate of the $s$-th node ($s<t$)~\cite{papadopoulos2012popularity}.  Thus, the hyperbolic distance $x_{st}$ mixes the effect of popularity, reflected by the radial distance, and similarity in the generic feature, quantified by the angular difference $\theta_{st}$. As shown in~\cite{papadopoulos2012popularity}, the models following the traditional preferential-attachment mechanism and the popular fitness model are naturally recovered for particular choices of the distributions of popularity $r_t$ and similarities $\theta_{st}$.

Given the accuracy in predicting the connection probability of technological and biological networks~\cite{papadopoulos2012popularity} and the ability of recovering other growth processes, the geometric popularity $\times$ similarity model has thus introduced a unifying framework for modeling network evolution in which vertex similarities can be taken into account~\cite{crandall2008feedback,ma2007modeling,watts2002identity,menczer2004correlated,javarone2013perception}.



\subsection{Time series}

Complex network theory is a suitable framework for studying any kind of complex system composed by discrete elements, whose interactions are described by the observed connectivity pattern. However, in many physical systems, these connections are not obviously revealed, requiring a further analysis based on the properties of the system in order to uncover the network topology. This is the case of networks constructed through time series analysis. Such networks are composed by nodes for which the accessible physical quantity is the time evolution of a certain property. Real-world examples that fit this definition are, for instance, financial market networks, whose nodes are assets with time evolving prices; cortical networks, constructed through functional brain analysis; climate networks, whose nodes are points in the globe with time evolving climate variables (e.g., temperature, pressure and humidity); and many others. The statistical similarities of time evolving quantities  are then the observable features that must be  taken into account for the inference of connections. In this section we briefly discuss some of the approaches to construct such networks in the light of the concepts presented in this review, i.e., how the time series associated to nodes (features) configure similarity spaces through which the connectivity pattern is generated.


\subsubsection{Financial market networks}

The seminal work by Mantegna~\cite{mantegna1999hierarchical} consists in one of the first approaches to treat a set of stocks as a network. Seeking to quantify the hierarchical organization of a portfolio of stocks, the author adopted as a similarity measure between pairs of stocks the correlation coefficient given by
\begin{equation}
\rho_{ij}=\frac{\left\langle Y_{i}Y_{j}\right\rangle -\left\langle Y_{i}\right\rangle \left\langle Y_{j}\right\rangle }{\sqrt{\left(\left\langle Y_{i}^{2}\right\rangle -\left\langle Y_{i}\right\rangle ^{2}\right)\left(\left\langle Y_{j}^{2}\right\rangle -\left\langle Y_{j}\right\rangle ^{2}\right)}},
\label{Eq:correlation_coefficient_rho}
\end{equation}
where $Y_i=\ln P_i(t) - \ln P_i(t-1)$ is the return and $P_i(t)$ the closing price at day $t$ of stock $i$. Since $\rho_{ij}$ does not fulfill the four axioms that define a metric (presented in Section~\ref{s:similarity}), Mantegna adopted instead
\begin{equation}
d(i,j) = \sqrt{2(1 - \rho_{ij})}.
\label{Eq:Mantegna_d_ij}
\end{equation}
It turns out
that Equation~\ref{Eq:Mantegna_d_ij} is simply the Euclidean distance between the stocks in the $N_t$-dimensional space in which the
$n$th-coordinate of stock $i$ is $(Y_i(n) - \left\langle Y_i \right\rangle)/\sigma^2_i$, where $\sigma_i$ is the standard deviation of
$Y_i$, $N$ the number of stocks and $N_t$ the number of negotiable days. Having defined the distance matrix $\mathbf{D}=[d_{ij}]$, the respective minimum spanning tree (MST) associated to the matrix can be calculated. The analysis of MST in financial market networks is a powerful
technique to identify clusters of companies in such systems, since it leads to the graph with the lowest cost in terms of the total distance required to create a path connecting all nodes. For instance, it was found that companies tend to form clusters according to their
economic sector~\cite{bonanno2003topology,garas2007correlation,kantar2012analysis,tumminello2007correlation}. Furthermore,
the MST framework allows the generalization to other metric spaces, which can be used to further analyze portfolios of
stocks~\cite{mantegna1999hierarchical} .


The framework introduced in~\cite{mantegna1999hierarchical} has been extensively studied and generalized in order to better understand financial systems. One important extension is the analysis of the time-dependent MST. For instance, by constructing the correlation matrix
using time-sliding windows, Onnela et al.~\cite{onnela2003dynamics} analysed topological measures of the originated networks as a function
of time.
The authors found that during
financial crashes the topological distances between the stocks tend to
decrease, as a result of the emergence of a strong correlation pattern
between the time series, shrinking the space in which the stocks are
embedded. Similar results were also reported in~\cite{fenn2011temporal,onnela2003dynamics,conlon2009cross}.

The analysis of financial markets using tools originated from network theory is a wide research field with many other applications, such as current exchange rate~\cite{fenn2009dynamic,eryiugit2009network,keskin2011topology}. We refer to~\cite{costa2011analyzing} for an overview of such applications.



\subsubsection{Climate networks}

Despite being relatively new in network analysis, the field of climate networks~\cite{tsonis2006networks,tsonis2008role,tsonis2008topology,donges2009backbone,donges2009complex,gozolchiani2008pattern,tsonis2004architecture,yamasaki2008climate} has been yielding important insights and results in climate sciences. The methodological approach of constructing networks through climate datasets is, in fact, similar to those applied in financial market networks, though with particular differences.
In general, nodes correspond to points in spatio-temporal grids over the globe, in which the links quantify statistical similarity between time series of climate variables associated to each node. In other words, in climate networks, nodes are already embedded in a well defined space and the similarity of time-evolution quantities are used to identify spatial patterns in order to relate them with climate effects. Examples of this can be found, for instance, in the identification of highly connected nodes associated to North Atlantic Oscillations~\cite{yamasaki2008climate,gozolchiani2008pattern,donner2008nonlinear,radebach2013disentangling,ludescher2013improved,ludescher2014very}, long range connections related to surface ocean currents~\cite{donges2009backbone,donges2009complex}, dense stripes of links in the tropics associated to Rossby Waves~\cite{wang2013dominant} and others~\cite{phillips2015graph}. Another important emergent pattern in climate networks is the so-called \textit{teleconnections}, edges that connect nodes separated by long geographical distances. These special links are non-trivial structures and constitute remarkable properties of such networks, since they act as shortcuts introducing small-world effects in the network~\cite{donges2009complex}.


Similarly to financial market networks, statistical similarities in climate data can also be quantified by the Pearson correlation coefficient in order to stablish the connections. In fact, most of the earlier works on the topic~\cite{tsonis2004architecture,tsonis2006networks,tsonis2008topology,yamasaki2008climate,gozolchiani2008pattern,donges2009backbone} were based on climate networks constructed through linear cross-correlation between time series associated to each grid-point in the climate data set. Furthermore, in order to avoid spurious effects in the analysis, it is extremely important that the time series have the natural seasonal effects removed, so that only the temporal anomalies are quantified. For time-series with high temporal resolution, the time-delayed Pearson correlation coefficient defined as
\begin{equation}
\rho_{i,j}^{(t)}(-\tau)=\frac{\left\langle T_{i}(t)T_{j}(t-\tau)\right\rangle -\left\langle T_{i}(t)\right\rangle \left\langle T_{j}(t-\tau)\right\rangle }{\sqrt{\left\langle \left(T_{i}(t)-\left\langle T_{i}(t)\right\rangle \right)^{2}\right\rangle \left\langle \left(T_{j}(t-\tau)-\left\langle T_{j}(t-\tau)\right\rangle \right)^{2}\right\rangle }}
\label{eq:Cij_mtau}
\end{equation}
and
\begin{equation}
\rho_{i,j}^{(t)}(\tau)=\frac{\left\langle T_{i}(t-\tau)T_{j}(t)\right\rangle -\left\langle T_{i}(t-\tau)\right\rangle \left\langle T_{j}(t)\right\rangle }{\sqrt{\left\langle \left(T_{i}(t-\tau)-\left\langle T_{i}(t-\tau)\right\rangle \right)^{2}\right\rangle \left\langle \left(T_{j}(t)-\left\langle T_{j}(t)\right\rangle \right)^{2}\right\rangle }}
\label{eq:Cij_tau}
\end{equation}
should be computed, where $T_i$ is the time-series associated to node $i$, $\tau$ is the time-delay and $\left\langle \cdot \right\rangle$ denote the time average over a given period. Studies on climate networks constructed through the coefficients in Equations~\ref{eq:Cij_mtau} and~\ref{eq:Cij_tau} have been yielding important results specially concerning the climate variability due to the El Ni\~{n}o Southern Oscillation (ENSO)~\cite{yamasaki2008climate,gozolchiani2008pattern,donner2008nonlinear,radebach2013disentangling,ludescher2013improved,ludescher2014very}.

Given the nonlinear fluctuations inherent in climate systems~\cite{phillips2015graph}, another adopted measure  to quantify the similarity and construct climate networks is the mutual information between the time series associated to given two nodes $i$ and $j$,
\begin{equation}
M_{ij}  = \sum_{mn} p_{ij}(m,n) \log \frac{p_{ij}(m,n) }{p_i(m) p_j(n)},
\label{Eq:Mutual_information_climate}
\end{equation}
with $p_i(m)$ and $p_{ij}(m,n)$ being, respectively, the marginal and joint probability density functions of the time series of given climate variables $x_i$ and $x_j$. The mutual information has the advantage of being able to quantify the relationship between two time series having strong nonlinear relationships. Moreover, networks constructed through mutual information analysis can reveal links that are absent in correlation-based structures, since nonlinear relationships between time series can lead to high values of $M_{ij}$, whereas such effects yield low values
of Pearson correlation coefficient~\cite{donges2009backbone}.

Many other measures to embed climate networks in feature-similarity space have been employed in order to track important climate events. For instance, phase coherence to detect relations between ENSO and the Indian Monsoon~\cite{maraun2005epochs},  entropy based on phase synchronization measures~\cite{yamasaki2009climate}, event-synchronization~\cite{malik2010spatial,malik2012analysis}  and also directed measures in order to detect causality in climate networks~\cite{ebert2012new,ebert2012causal,runge2012escaping,runge2012quantifying}.

\subsubsection{Functional brain networks}

The brain represents a high metabolic cost for the body. Therefore, the vast network of connections between neurons and brain modules need to be as efficient as possible, that is, produce a high processing power with a small cost. This cost increases with neuronal density, as well as axonal density, diameter and length \cite{bullmore_economy_2012}. In addition, the connectivity also needs to be robust against perturbations. Therefore, unveiling the mechanisms underlying such optimal connectivity between neurons and brain regions is an important topic \cite{cuntz_one_2010,chklovskii_exact_2004}.

There are two main networked systems of interest in the brain, the structural one and the functional one. The structural system is the network formed by physical connections between neurons, or in a more coarse grained view, the so-called neural pathways between brain modules. The functional system is formed by the dynamics of information exchange between processing regions of the brain. In \cite{honey_predicting_2009} diffusion spectrum imaging (DSI) and functional magnetic resonance imaging (fMRI) was used to obtain, respectively, the structural and resting state functional connectivity of 998 cortical regions. The authors found that the presence of strong structural connectivity between two regions is a good indicator of a strong resting state functional connectivity, but regions without structural connectivity could still present strong resting state functional connectivity. This means that inference of structural connectivity from functional connectivity is not reliable. Nevertheless, the study of the networks formed by information exchange is still of great interest, since in many aspects the actual dynamics evoked in the brain may be considered more relevant than its underlying structure.

In order to construct a functional brain network, one begins by measuring the activity of brain modules along time, this defines a set of time series that can be compared to produce the network. There are many methods in the literature to verify if two time series are related, a comprehensive analysis of such methods was made by Smith et al. \cite{smith_network_2011}. They used a rigorous framework to generate simulated functional magnetic resonance imaging (fMRI) time series in known network topologies, and compared the efficiency of many measurements to uncover which nodes were connected in the network. The methods were divided into two categories, the first being methods used to find undirected connections, that is, with no prediction of causality, and the second being methods that can unveil causality between the time series. In Table~\ref{t:fmri_methods} we indicate the methods considered by the authors. They found that the top performing methods for non-causal connectivity prediction when considering all simulation conditions were partial correlation, regularised inverse covariance and Bayes net methods. But it is important to observe that, with the exception of one single experiment, the networks used in the simulations were all directed with zero reciprocity. The only experiment having nonzero reciprocity resulted in a similar accuracy between most of the methods. One striking result of the study is that all causality prediction methods performed poorly on the experiments.

\begin{table}[htb]
\tbl{\label{t:fmri_methods} Methods tested in~\protect\cite{smith_network_2011} for estimating functional brain networks. We note that LiNGAM stands for Linear Non-Gaussian Acyclic Model.}
{\begin{tabular}{lc}
  \hline
  {\bf Name} & {\bf Key reference} \\
  \hline
  \hline
  Pearson correlation & \cite{snedegor1967statistical} \\
  Partial correlation & \cite{marrelec2006partial} \\
  Regularised inverse covariance & \cite{banerjee2006convex} \\
  Mutual information & \cite{cover2012elements} \\
  Generalized synchronization & \cite{quiroga2002performance} \\
  Patel's conditional dependence & \cite{patel2006bayesian} \\
  Wavelet transform coherence & \cite{torrence1998practical} \\
  Bayes net methods & \cite{spirtes2000causation} \\
  LiNGAM & \cite{shimizu2006linear} \\
  Granger causality & \cite{granger1969investigating} \\
  Partial directed coherence & \cite{baccala2001partial} \\
  Directed transfer function & \cite{kaminski1991new} \\
  \hline
\end{tabular}
}
\end{table}

\section{Metric embedding}

Metric embedding involves finding an appropriate feature space in which the \emph{distances} between objects provide a good description of the known \emph{dissimilarity} between them~\cite{indyk2004low,abraham2006advances}. Such a procedure represents the SF path indicated in Figure~\ref{f:all_paths}. The feature space is usually considered to be Euclidean, since one usually seeks a more intuitive representation of the system when doing metric embedding. The concept of dissimilarity, or similarity, is used here in a broad sense, as it does not necessarily need to follow any constraints such as triangular inequality or nonnegativity. One of the most well-known techniques for embedding a similarity matrix is multidimensional scaling~\cite{borg2005modern}, which will be the focus of the discussion in this section. Nevertheless, another common technique used for metric embedding is the Isomap~\cite{tenenbaum2000global}, which uses information about the neighborhood of each point to construct an appropriate manifold for embedding.

\subsection{Multidimensional scaling}

Multidimensional scaling (MDS) is a powerful technique able to find the positions of objects in a feature space when only the similarities between the objects are known. One of the most traditional uses of the method is to allow the visualization of entities according to human perception \cite{liu2004measurement,wish1970differences,garner2014processing,borg1983dimensional}. By evaluating a set of entities according to their similarities, researchers are able to construct a robust psychological map of a given concept \cite{borg2005modern}. When the values of the original features of the objects are known, MDS is commonly used as a dimensionality reduction technique \cite{webb2003statistical}. In such a case, the data is transformed from the feature space to the similarity space and projected back to the feature space, following a FSF path.

\begin{figure}[!htbp]
\begin{center}
\includegraphics[width=0.6\linewidth]{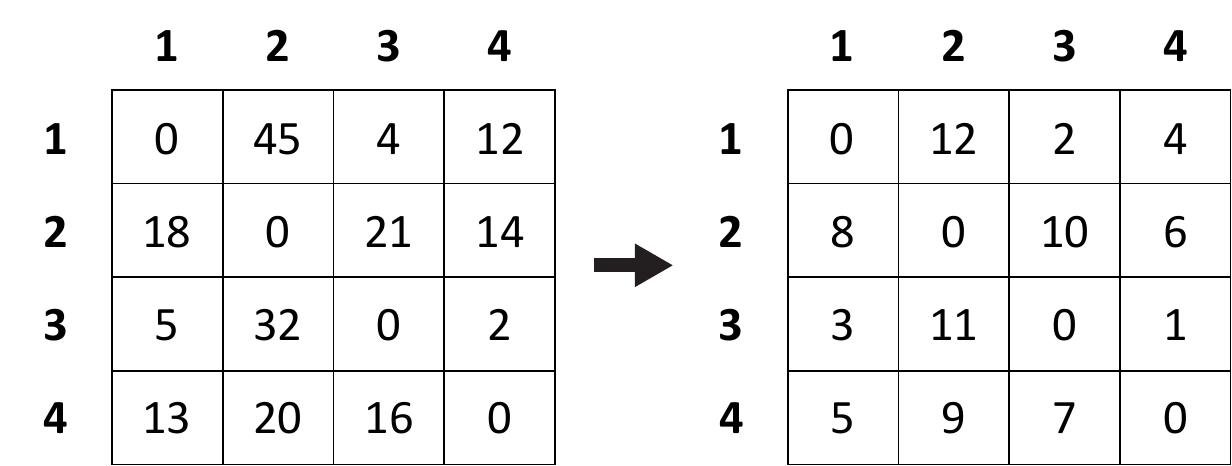}
\end{center}
\caption{Example of transforming a 4x4 similarity matrix (left) into a ranking matrix (right).}
\label{t:OMDS}
\end{figure}

The main input of any MDS algorithm is the similarity matrix, $\mathbf{S}$, between the objects and the specific MDS algorithm used depends on the properties of this matrix. Since the method is commonly used to embed the points into an Euclidean space, a process which will ultimately produce an Euclidean distance matrix $\mathbf{D}$ between the points, the method used for MDS depends on how close to a Euclidean distance matrix the matrix $\mathbf{S}$ is. The most basic form of MDS, called spectral MDS (SMDS), is used when the similarity matrix is expected to be very close to an Euclidean distance matrix. Note that in this case $\mathbf{S}$ is actually a dissimilarity matrix (i.e., objects that are similar have lower value in the matrix). Strikingly, this method is known to work well even in some cases where the similarity matrix is far from being an Euclidean distance matrix \cite{webb2003statistical}. Another class of methods for multidimensional scaling are optimization methods. Such methods are commonly divided in two main classes \cite{borg2005modern,borg2012applied} a) metric and b) non-metric. Metric multidimensional scaling (MMDS) is applied when the mapping between the similarity matrix and the resultant Euclidean distance matrix is a well-defined function $f$.  Non-metric multidimensional scaling (NMMDS) is used when the only restriction on $f$ is that the function must be monotonic. The most usual form of NMMDS is called ordinal multidimensional scaling (OMDS), where the values of the similarity matrix are transformed into rankings. Refer to Figure~\ref{t:OMDS} for an example of such transformation. 

As said above, the purpose of any MDS algorithm is to find a configuration of points where the elements of the similarity matrix $s_{ij}$ are as close as possible to the elements of the Euclidean distance matrix $d_{ij}$ of the points placed into the embedding space. The goodness-of-fit of such a relationship can be measured in different ways, the most common one being the definition due to Kruskal \cite{kruskal1964multidimensional}, called Stress-1, and expressed by

\begin{equation}
\sigma_1 = \sqrt{\frac{\sum_{i<j} (s_{ij}-d_{ij})^2}{\sum_{i<j} d_{ij}^2}}.
\end{equation}
Another commonly studied property is the \emph{stress per point}, which is the stress evaluated for a single point

\begin{equation}
\sigma_p(i) = \sqrt{\frac{\sum_j (s_{ij}-d_{ij})^2}{\sum_{ij} d_{ij}^2}}.
\end{equation}
This property can be used to indicate how well the distances are being represented for a particular point. This in turn can be used to study which points would need more dimensions to be better represented in the projection, that is, it can give information of the local dimensionality of the data. It should be pointed out that currently there is no universal method to assess what is a ``good" value for stress. It is known that stress is influenced by the number of points, dimension of the projection, number of proximity ties, noisy or missing data and also on the relationship between $s_{ij}$ and $d_{ij}$ \cite{borg2005modern}.


\subsection{Multidimensional scaling on graphs}
\label{sec:mult_scaling_graphs}

Multidimensional scaling can also be used to visualize graphs. In such a case, the procedure represents the path in Figure~\ref{f:all_paths} departing from the connectivity, passing trough the similarity between nodes and arriving at adequate features to visualize the network (path CSF). In \cite{lee_embedding_2012} non-metric multidimensional scaling was used to compare the embedding of real-world networks with Erd\H{o}s-R\'enyi \cite{erdos1960evolution}, Barab\'asi-Albert \cite{barabasi1999emergence} and Watts-Strogatz \cite{watts1998collective} networks. Two similarity measurements were used, structural equivalence \cite{lorrain_structural_1971}, which measures the number of similar nodes in the neighborhood of two nodes, and the sum of edge betweenness weights of the shortest path between two nodes. An interesting result was that the network models usually required more dimensions to be correctly represented than real-world networks. In \cite{toivonen_networks_2012} the authors analyzed relationships between words describing emotional experiences, and compared the conclusions that can be drawn from a purely MDS study with characterizations obtained from traditional network measurements. Multidimensional scaling has also been used for the visualization of the connectivity between brain regions \cite{scannell_connectional_1999,salvador_neurophysiological_2005,stephan_computational_2000}, in order to reveal relationships or hierarchies between such regions. Another useful application of the technique on graphs is to visualize effective travel time between geographical locations \cite{viana2011fast}.

In the following we will briefly describe a process that allows the visualization of a graph using MDS. First, a common strategy \cite{gansner2005graph} is to define the stress function

\begin{equation}
\sigma_G(X) = \sum\limits_{i=1}^{N-1}\sum\limits_{j=i+1}^{N} w_{ij}(s_{ij}-d_{ij})^2 \label{eq:stress_g}
\end{equation}
and choose appropriate values for the similarities $s_{ij}$ and weights $w_{ij}$ according to topological properties of the graph. The well-known graph visualization software Graphviz~\footnote{http://www.graphviz.org/} defines $s_{ij}$ as the shortest path length between nodes $i$ and $j$ \cite{gansner2005graph}, which we call $l_{ij}$. The weights are chosen in a way that topologically closer nodes are more important to the stress function. A good choice of $w_{ij}$ was empirically found to be $w_{ij}=1/l_{ij}^2$~\cite{gansner2005graph}. 

Let the $N\times d$ matrix $X$ represent the position of the $N$ nodes in a $d$-dimensional space, the stress defined in Equation \ref{eq:stress_g} can then be expressed as

\begin{equation}
\sigma_G(X) = \sum\limits_{i=1}^{N-1}\sum\limits_{j=i+1}^{N} w_{ij} (l_{ij}-||X_i-X_j||)^2 \label{eq:stress_g2}
\end{equation}
where $||X_i-X_j||=d_{ij}$. Note that the inclusion of a weight proportional to $1/l_{ij}^\alpha$, which is a common practice in the literature, means that topologically closer nodes have more restricted relative positions. Also, if $w_{ij}=1/l_{ij}^2$ expression \ref{eq:stress_g2} is identical to the expression used by Kamada and Kawai \cite{kamada1989algorithm} for their well known spring-embedded graph visualization algorithm. But contrary to the expression defined in \cite{kamada1989algorithm}, one can use distinct topological properties for $s_{ij}$ and $w_{ij}$ in Equation \ref{eq:stress_g} and define different appropriate stress functions \cite{cohen1997drawing}. These can in turn be used to provide distinct projections for the same graph, which may bring new insights about its structure. Nevertheless, a fundamental difference between the two methods is that while the Kamada-Kawai method uses a Newton-Raphson algorithm to minimize the energy of the system, multidimensional scaling naturally uses a technique called stress majorization. The latter is known for providing faster convergence and less chance of being stuck in a local minimum \cite{gansner2005graph}. 

The stress majorization can be easily applied to a graph. First, the nodes are placed in a $d$-dimensional space, the position of the nodes can be randomly drawn or they may reflect some property of the graph (e.g., the presence of communities). Then, the following \emph{majorizing function} is defined:

\begin{equation}
F(X,Z) = \sum\limits_{i<j} w_{ij}l_{ij}^2 + Tr(X^T L^w X) - 2Tr(X^T Q Z)\label{eq:maj_func_F}
\end{equation}
where $Z$ is the $N\times d$ matrix containing the current position of the nodes, $L^w$ is the weighted Laplacian of the network

\begin{equation}
L_{ij}^w = 
\begin{cases}
	-w_{ij} & \text{if  } i\neq j \\
	\sum_{k\neq i} w_{ik} & \text{if  } i=j
\end{cases}
\end{equation}
and $Q$ is the matrix

\begin{equation}
Q_{ij} = 
\begin{cases}
	-w_{ij}l_{ij}\text{inv}(||Z_i-Z_j||) & \text{if  } i\neq j \\
	\sum_{k\neq i} w_{ik}l_{ik}\text{inv}(||Z_i-Z_k||) & \text{if  } i=j
\end{cases}.\label{eq:mds_Q}
\end{equation}
In Equation~\ref{eq:mds_Q}, $\text{inv}(x)=1/x$ when $x\neq 0$ and 0 otherwise. Also, note that in Equation~\ref{eq:maj_func_F} we have $F(Z,Z)=\sigma_G(Z)$. The minimum of the majorizing function can be found by solving for X

\begin{equation}
L^w X = Q Z. \label{eq:iter}
\end{equation}
Iteratively minimizing the majorizing function is equivalent to minimizing the stress of the projection, given by Equation \ref{eq:stress_g2}. This means that in order to find appropriate positions for the nodes in the graph, it suffices to solve 

\begin{equation}
L^w X(t+1) = Q X(t). \label{eq:iter2}
\end{equation}
Note that matrix $Q$ needs to be recalculated at each step, considering $Z=X(t)$ for the calculation. Since $L^w$ is not of full rank, it does not have an inverse. One method to solve Equation \ref{eq:iter2} is to calculate the Moore-Penrose pseudoinverse of $L^w$ \cite{ben2003generalized}. A faster solution indicated in \cite{gansner2005graph} is to solve the system of equations by using Cholesky factorization or conjugate gradient.

\begin{figure}[!htbp]
\begin{center}
\includegraphics[width=0.9\linewidth]{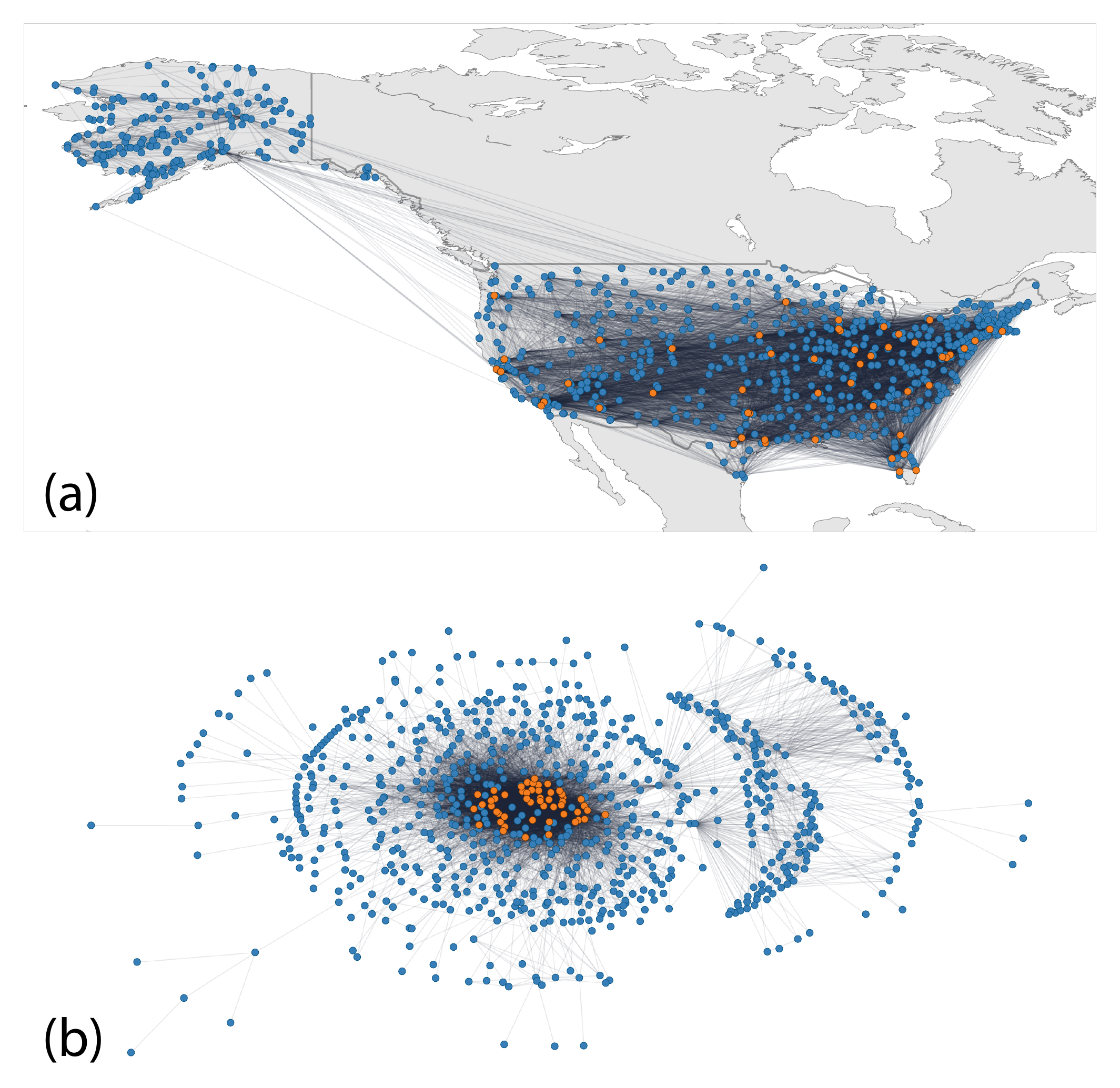}
\end{center}
\caption{Illustration of the structural information captured by the multidimensional scaling (MDS) technique. (a) United States airport network, where nodes represent airports and they are connected if there was a flight between them in the year 2013. The most important airports (see text for definition) are colored in orange. (b) Result of the MDS algorithm applied on the airport network, where the number of passengers flying between airports was considered as the similarity measurement.}
\label{fig:MDS_airport}
\end{figure}

In order to illustrate the methodology described above, we apply it to an airport network. We use the dataset provided by the United States Department of Transportation\footnote{Available at http://www.transtats.bts.gov/DataIndex.asp}, which contains information about airline routes between United States airports. The dataset also contains the position of each airport and the number of passengers traveling each month between airports. The airline routes are used to construct a graph, where nodes represent airports and two nodes are connected if the respective airports shared a flight in the year 2013. We used the real positions of the airports to visualize the graph on a map of the United States, as can be seen in Figure \ref{fig:MDS_airport}(a). Nodes colored in orange represent airports through which more than $5\times 10^5$ passengers traveled  in the year of 2013, therefore, they can be considered as the most important nodes in the network. Each connection can be associated with the number of passengers that traveled between the two airports connected by the edge in the year 2013. We used this information to construct a MDS projection of the airport network, as shown in Figure \ref{fig:MDS_airport}(b). The points have the same color as in Figure \ref{fig:MDS_airport}(a). We see that the most important airports define the core of the network. They, in turn, are surrounded by secondary airports, while smaller airports are placed far from the core, scattered throughout the space. We also see that the embedding preserved the presence of the community composed of Alaskan airports. The MDS embedding can also be used for additional analysis, such as coloring the nodes according to their topological properties or visualizing a dynamics applied to the network, such as cascading failures \cite{albert2000error,motter2002cascade}.

\section{Selection}
\label{s:selection}

The selection transformation concerns choosing which edges from a similarity matrix should be considered as valid relationships. The most common procedure to eliminate spurious relationships is to apply a simple threshold to the matrix. That is, values smaller than the threshold become zero and those that are larger than the threshold define the connections of the respective graph. Nevertheless, many other options exist for applying the selection, such options are usually tied to the placement of the objects in the embedding space. Thus, many traditional geographic network models can be related to the selection procedure. We present below models that are commonly used in the literature, as well as some models that introduce new concepts related to this procedure.

\subsection{Random geometric graphs}
\label{s:geometric_graphs}

The random geometric graph model is one of the simplest approaches to define a spatial network. Given a set of points placed in a space, each point is connected if they are separated by a distance smaller than $R$. Equivalently, one can imagine that the points are replaced by disks (spheres or hyperspheres if in more than 2 dimensions) with radius $R$ and, when two disks overlap, the respective points are connected by a link. This procedure defines a feature to similarity transformation, and another transformation from similarity to connectivity (path FSC). Random geometric graphs can be used to model a range of systems, including wireless sensor networks~\cite{akyildiz2002wireless,yick2008wireless}, ad hoc networks~\cite{nemeth_giant_2003}, contact networks by using mobile agents~\cite{gonzalez_system_2006,gonzalez_scaling_2004} and even clustering of data~\cite{penrose_random_2003}. Besides the parameter $R$, the position of the points also influences on the generated topology. Such positions are drawn according to some given distribution $p(\mathbf{r})$. Usually, it is assumed that $p(\mathbf{r})$ is a uniform distribution, that is, $p(\mathbf{r})=p_0$. In this case, the mean degree of the network can be written as~\cite{dall_random_2002}

\begin{equation}
\langle k\rangle = \frac{(R\sqrt{\pi})^d N}{\Gamma(\frac{d+2}{2})},
\end{equation}
where $d$ is the dimension of the embedding space. By letting $N\rightarrow\infty$ and $R\rightarrow 0$, Herrmann et al.~\cite{herrmann_connectivity_2003} showed that the degree distribution of a random geometric graph generated from points distributed according to $p(\mathbf{r})$ is

\begin{equation}
P(k,\alpha)=\frac{\alpha ^k}{k!}\int\!p(\mathbf{r})^{k+1}e^{-\alpha p(\mathbf{r})}\,\mathrm{d}\mathbf{r},
\label{eq:degree_dist_geo}
\end{equation}
where $\alpha = \langle k\rangle /\int\!{p(\mathbf{r})^2 \mathrm{d}\mathbf{r}}$. Note that if $p(\mathbf{r})=p_0$, the degree distribution of the network reduces to the Poisson distribution, which is also found for the Erd\H{o}s-R\'enyi model. Despite such similarity between the degree distribution of random geometric and Erd\H{o}s-R\'enyi graphs, they are sharply distinct models. This happens because, in contrast to the Erd\H{o}s-R\'enyi model, geometric graphs do not have independence of edge existence, that is, two nodes connected to a common third node are likely to be connected between themselves. 

\subsection{Planar graphs}

The idea of a planar graph model is to connect points in a plane so that there are no edge crossings \cite{trudeau2013introduction}. Examples of real world quasi-planar networks include the electric grid~\cite{sole2008robustness}, streets~\cite{lammer2006scaling} (with just some occasional non-planar edges as a result of an overpass or tunnel) and the hallways in an exhibition. In 1996 the idea of a random planar graph was introduced~\cite{denise_random_1996} and the strong conceptual similarity with Erd\H{o}s-R\'enyi graphs has led some authors to call it planar Erd\H{o}s-R\'enyi graphs~\cite{barthelemy2011spatial}. 

The procedure to generate a random planar graph with $N$ nodes is as follows~\cite{denise_random_1996}. Starting from any initial simple graph with $N$ nodes (usually the empty graph) a pair of nodes $(i,j)$ is randomly drawn with equal probability for all pairs. If the pair is already connected, the edge is deleted. If they are not connected, a new edge is added between the pair if the graph remains planar after adding the edge. With such procedure it is possible to define an irreducible aperiodic Markov process having a symmetric transition matrix. The stationary distribution of the process samples with equal probability the set of planar graphs having $N$ nodes. In practical terms, one just needs to repeat the process of randomly drawing pairs of nodes for a sufficiently long time, which depends on the number of nodes and initial graph structure. Note that the similarity between nodes is never considered directly. Still, nodes that are close to each other are more likely to be connected since they tend to maintain the planarity of the network.

We note that, although random planar graphs have many interesting theoretical implications, the structure generated by these graphs has hardly any visual resemblance with real-word networks, with some rare exceptions~\cite{masucci_random_2009}. Nevertheless, a growth model with some additional constraints that can generate planar networks having some similarities with urban street networks have been defined~\cite{barthelemy_modeling_2008}.

\subsection{Spatial small-worlds}

The Watts-Strogatz (WS) model considers as initial network a ring lattice and the links are randomly rewired with a probability $p$ ~\cite{watts1998collective}. Generalizations of the WS model to spatially embedded systems treat the probability of rewiring two nodes as depending on the distance between them, since it is assumed that longer shortcuts are expected to have higher costs in real spatial networks~\cite{barthelemy2011spatial}. Starting from regular lattices with $N$ nodes embedded in a $d$-dimensional space, shortcuts are typically added with probability given by
\begin{equation}
P(\ell) \sim \ell^{-\alpha}, 
\label{eq:p_sw_spatial}
\end{equation} 
where $\ell$ is the distance between two nodes in the lattice and $\alpha \geq 0$~\cite{kleinberg2000navigation,jespersen2000small,sen2001small,petermann2006physical}. 
As remarked in~\cite{petermann2006physical}, the inclusion of spatial constraints when adding shortcuts can yield interesting properties when compared with the original small-world (SW) model. For instance, the performance of dynamical processes such as navigability, random walks and diffusion will strongly depend on the exponent $\alpha$. In fact, as conjectured in~\cite{kasturirangan1999multiple}, the small-world phenomenon is due to the emergence of multiple length scales, which is in agreement with Equation~\ref{eq:p_sw_spatial}~\cite{petermann2006physical}. 

Similarly as in the original formulation, it is expected that the geographic generalizations of the WS model also present a transition between large- and small-world regimes. Since the small-world regime is characterized by low values of averaged shortest path length and high values of clustering coefficient, we also expect that the crossover between large- and small-world regime will take place at a certain value for the exponent $\alpha_c$. 
As shown in~\cite{petermann2006physical}, the average distance $\left\langle \ell \right\rangle$ for regular networks embedded in a $d$-dimensional space with shortcuts added with probability given by Equation~\ref{eq:p_sw_spatial} follows
\begin{equation}
\left\langle \ell \right\rangle = L^* \mathcal{F}_\alpha \left( \frac{L}{L^ *}\right),
\label{eq:scaling_distance_sw}
\end{equation}
where the characteristic length $L^*$  scales with
\begin{equation}
L^*(p)\sim\begin{cases}
p^{-1/d} & \textrm{ if }\alpha < 1\\
p^{1/(\alpha - d - 1)} & \textrm{ if }\alpha > 1
\end{cases}
\label{eq:scaling_function_distance_sw}
\end{equation}
and the scaling function $\mathcal{F}_\alpha$ obeys
\begin{equation}
\mathcal{F}_{\alpha}(x)\sim\begin{cases}
x & \textrm{ if }x\ll1\\
\ln x & \textrm{ if }x\gg1. 
\end{cases}
\label{eq:scaling_function_distance_sw2}
\end{equation}
Thus, the transition between the large- and small-world is characterized by the threshold exponent $\alpha_c = d + 1$.

Other models with similar properties as in~\cite{petermann2006physical} were proposed in~\cite{jespersen2000small,sen2002phase,moukarzel2002shortest}, in which it was also observed
a threshold $\alpha_c$ characterizing the transition between the regimes of large- and small-world depending on the system's dimension. 

\subsection{Spatial scale-free model}
\label{sec:spa_scale_free}

In the Babar\'asi-Albert (BA) model, the mechanism that leads to power-law degree distribution is the so-called preferential attachment, in which at each step of the process a new node $j$ is created and connects to other $m$ nodes already present with probability
\begin{equation}
P_{j\rightarrow  i } \propto k_i, 
\label{eq:pij_ba}
\end{equation}
where $k_i$ is the degree of node $i$. Many spatial growth models consider a combination of preferential attachment and distance to define the connectivity~\cite{yook2002modeling,rozenfeld2002scale,warren2002geography,sen2002phase,jost2002evolving,manna2002modulated,xulvi2002evolving,barthelemy2003crossover}. Typically, the probability that a new node $j$ will connect to other nodes in the network is given by~\cite{barthelemy2011spatial}
\begin{equation}
P_{j \rightarrow i } \propto k_i \mathcal{F}[d_{ij} ],
\label{eq:pij_ba_spatial}
\end{equation} 
with $\mathcal{F}$ being a function of the Euclidean distance $d_{ij}$ between nodes $i$ and $j$. For instance, Barth\'elemy~\cite{barthelemy2003crossover} explored the following model. First, $N$ nodes are randomly distributed in a $d$-dimensional space having linear size $L$. Then, nodes are connected with probability given by Equation~\ref{eq:pij_ba_spatial}, but setting $\mathcal{F}(d_{ij}) = e^{-d_{ij}/r_c }$ for the distance function, where $r_c$ is a finite scale. For sufficiently large values of $r_c$ the model behaves as the traditional BA model and exhibits scale-free degree distributions with no influence of the embedding space. For $r_c \ll L$ it can be shown that the degree distribution is given by~\cite{barthelemy2003crossover}
\begin{equation}
P_k = k^{-\gamma}f\left(\frac{k}{k_c}\right), 
\label{eq:p_k_ba_spatial}
\end{equation}
where $\gamma  = 3$ and $f$ is a scaling function with cutoff $k_c \sim n^\beta$, where $\beta = 0.13$ and $n$ is the average number of points in a sphere of radius $r_c$, given by~\cite{barthelemy2003crossover}
\begin{equation}
n = \rho r_c^d \frac{\pi^{d/2}}{\Gamma(1 + \frac{d}{2})}.
\label{eq:n_ba_spatial}
\end{equation} 

The model described above generates a power-law degree-distribution through the preferential attachment mechanism,  with the nodes being distributed uniformly in the embedded space. However, models that do not follow the preferential attachment paradigm can also lead to spatial scale-free networks through a proper placement of the points in space, showing that the nodes spatial distribution plays an important role in network connectivity~\cite{barthelemy2011spatial}. In particular, as discussed in Section~\ref{s:geometric_graphs}, depending on the chosen spatial distribution $p(\mathbf{r})$, the model proposed by Herrmann et al.~\cite{herrmann_connectivity_2003} produces different classes of networks. For instance, considering a one-dimensional model, in which the nodes are distributed in the interval $x \in [0,1]$ according to
\begin{equation}
p(x) = (1 - \beta)x^{-\beta};
\label{eq:p_x_herrmann_sf}
\end{equation}
the degree distribution in Equation \ref{eq:degree_dist_geo} is reduced to ~\cite{herrmann_connectivity_2003}
\begin{equation}
P(k;\alpha) \sim \frac{1}{\alpha \beta} [\alpha (1-\beta)]^{1\beta} k^{-1/\beta},
\label{eq:P_k_hermmann_sf}
\end{equation}
where $\beta < 1$ and $\alpha = \left\langle k \right\rangle /\int p^2(x) dx$. 

Preferential attachment and power-law distribution of nodes in space are not the only mechanisms to obtain graphs with power-law degree distributions. As shown by Bogu\~{n}\'a et al.~\cite{krioukov2010hyperbolic}, scale-free networks can also be naturally obtained on hyperbolic spaces. A network composed of $N$ nodes randomly placed in the two-dimensional hyperbolic space over a disk of radius $R$ has Euclidean radial distribution given by 
\begin{equation}
p(r) = \frac{\sinh r}{\cosh R - 1} \sim e^r.
\label{eq:density_hyperbolic_space}
\end{equation}
Connecting every pair of nodes $i$ and $j$ with probability $P(d_h(i,j))=\Theta(R - d_h(i,j))$, where $d_h(i,j)$ is the distance between $i$ and $j$ in the hyperbolic space and $\Theta(\cdot)$ the Heaviside function, it can be shown that the resulted degree distribution is given by
\begin{equation}
P(k)=2\left(\frac{\left\langle k\right\rangle }{2}\right)^{2}\frac{\Gamma(k-2,\left\langle k\right\rangle /2)}{k!}\approx k^{-3},
\label{eq:P_k_hyperbolic}
\end{equation}
where $\Gamma$ is the Gamma function.

\section{Topological similarity}
\label{sec:local_similarity}

The hidden metric space in which a network is embedded plays an important role in the observed topology of connections. Therefore, we expect nodes that are close in such metric space to have similar topological characteristics. Topological similarity measurements aim at uncovering the hidden metric defining the network connectivity. For example, it is possible to quantify, in a social network, the similarities between the nodes according to the cardinality of the set of common friendships and interests~\cite{missing}. Another example is the set of measurements devised to compute the similarity between two pieces of texts in citation networks~\cite{amancio2012using,Amancio2012427,menczer2004evolution,maria,hoax}. In this section, we focus on similarity measurements based on the topological structure of the networks. We note that calculating the topological similarity of nodes is a fundamental step in the link prediction problem~\cite{lu2011link,liben2007link}, which constitutes a CSC path. In this section, we classify the similarity in measures into two distinct groups: those based on local and global information of the network topology.

\subsection{Local similarity}

Local similarity measurements rely upon local information alone, i.e. the information of neighbors, neighbors of neigbhors and further hierarchies.
The simplest idea for computing the similarity considers that two nodes are similar whenever they share many neighbors. This approach is oftentimes referred to as \emph{structural similarity}~\cite{newman2010networks} and it is based on the assumption that the network topology already reflects a hidden information about the nodes. In terms of the adjacency matrix, the number of neighbors $q_{ij}$ shared by nodes $i$ and $j$ is given by
\begin{equation} \label{simples}
    q_{ij} = \sum_k a_{ik} a_{kj}.
\end{equation}
Equivalently, $q_{ij} = [\mathbf{A}^2]_{ij}$. Note that, according to Equation \ref{simples},
pairs of nodes with high degrees usually  share more neighbors than pairs of low-connected
nodes. To avoid this bias towards highly connected nodes, some kind of normalization is required.
A common normalizing factor is given by the geometric mean $\sqrt{k_i k_j}$, which leads to a modified version of $q_{ij}$, written as
\begin{equation} \label{quasi-cos}
    c_{ij} = \frac{q_{ij}}{\sqrt{k_i k_j}} = (k_i k_j)^{-\frac{1}{2}} \sum_k a_{ik} a_{kj}.
\end{equation}
Considering an unweighted and undirected network, we have that
\begin{equation}
 k_i = \sum_j a_{ij} = \sum_j a^2_{ij}.
\end{equation}
Hence, Equation \ref{quasi-cos} can be rewritten as
\begin{equation} \label{cosseno}
    c_{ij} = \frac{ \sum_k a_{ik} a_{kj} }{\sqrt{\sum_j a^2_{ij}}\sqrt{\sum_i a^2_{ij}}}.
\end{equation}
If the $i$-th and $j$-th rows of $\mathbf{A}$ are respectively represented as the vectors $\mathbf{a}_i$ and $\mathbf{a}_j$, then $c_{ij}$ can be seen as the cosine similarity, i.e.  the cosine of
the angle $\theta$ between $\mathbf{a}_i$ and $\mathbf{a}_j$, given by
\begin{equation} \label{cosseno}
    c_{ij} = \cos \theta = \frac{\mathbf{a}_i \cdot \mathbf{a}_j}{|\mathbf{a}_i||\mathbf{a}_j|}.
\end{equation}
Therefore, the similarity $c_{ij}$ ranges in the interval $[0,1]$.
This measurement has been employed, for example, to uncover the community structure of complex networks~\cite{yang}.

In addition to the geometric mean, other quantities have been used to normalize $q_{ij}$.
For example, a related normalization factor relying on node degrees, given by
\begin{equation}
    c_{ij} = q_{ij} / \min \{ k_i, k_j \},
\end{equation}
was employed to compute the overlap between substracts in the \emph{Eschericia Coli} metabolic network~\cite{Ravasz30082002}. Other simple normalizations include the Jaccard Index, the Sorensen Index and the Hub depressed index, given respectively by
\begin{equation}
    c_{ij}^{\textrm{J}} = \frac{q_{ij}}{| \Gamma_i \cup \Gamma_j |},
\end{equation}
\begin{equation} \label{sor}
    c_{ij}^{\textrm{S}} = \frac{2q_{ij}}{k_i + k_j},
\end{equation}
\begin{equation}
    c_{ij}^{\textrm{H}} = \frac{q_{ij}}{\max \{ k_i,k_j \} },
\end{equation}
where $\Gamma_i$ represents the set of neighbors of node $i$.

Another common normalization for $q_{ij}$ considers the expected number of shared neighbors in a null model of the network~\cite{newman2010networks}. If node $i$ picks each of its $k_i$ neighbors just by chance, then the likelihood for a given edge of $i$ to link to a neighbor of $j$ is $k_j/N$. After the random selection of $k_i$ neighbors, the expected number of shared neighbors will be $q_{ij}^{\textrm{(rand)}} = k_i k_j / N$. The similarity normalized by $q_{ij}^{\textrm{(rand)}}$ is then given by
\begin{align} \label{quasi-cov}
    q_{ij}^{\textrm{(norm)}}& =  q_{ij} - \frac{k_i k_j}{N} \nonumber \\
    & = \sum_k ( a_{ik} - \langle \mathbf{a}_i \rangle )( a_{jk} - \langle \mathbf{a}_j \rangle ),
\end{align}
Note that Equation \ref{quasi-cov} can be regarded as a covariance between $\mathbf{a}_i$ and $\mathbf{a}_j$. Such covariance can be normalized by the respective standard deviations of $\mathbf{a}_i$ and $\mathbf{a}_j$, which gives rise to the definition of the Pearson correlation
\begin{equation} \label{percor}
    c_{ij} = \frac{\sum_k ( a_{ik} - \langle \mathbf{a}_i \rangle )( a_{jk} - \langle \mathbf{a}_j \rangle ) }{ \sqrt{\sum_k ( a_{ik} - \langle \mathbf{a}_i \rangle ) ^ 2} \sqrt{( a_{jk} - \langle \mathbf{a}_j \rangle ^ 2)}}.
\end{equation}
%

Another well-known dissimilarity measurement based on the number of shared neighbors can be written in terms of the Euclidean distance.
Usually, this distance is normalized by the maximum distance between two vectors, i.e. $k_i + k_j$. Therefore, such a measurement is given by
\begin{align} \label{euclidiana}
    d(i,j) & = {\sum_k ( a_{ij} - a_{jk} ) ^ 2 }\Big{[} k_i + k_j \Big{]}^{-1} \nonumber \\
    & = 1 - 2 \frac{ q_{ij} }{k_i + k_j}.
\end{align}
The Euclidean distance in Equation \ref{euclidiana} is equivalent to the similarity measurement defined in Equation \ref{sor}.
Therefore, $d(i,j)$ is purely an alternative normalization for $q_{ij}$.

Some measurements based on neighbors have been inspired on concepts from language modeling~\cite{Ponte:1998:LMA}. For example, to overcome the problem of unseen bigrams~\cite{Keller:2003:UWO} (i.e. pairs of words that appear on the training test but do not occur on the test set), the words most similar to the unseen word compounding the bigram is chosen for a specific task~\cite{Keller:2003:UWO}. Analogously, this idea might be extended to compute node similarities. Suppose we are given the set $\mathcal{S}_i^{(k)}$, i.e. the set of the $k$-most related nodes to node $i$ according to a given similarity measurement. Then, the new similarity measure can be calculated as
\begin{equation}
    c_{ij}^{\textrm{N}} = | \{ z : z \in \Gamma(j) \cap \mathcal{S}_i^{(k)} \} |.
\end{equation}
%

Finally, some local approaches use the local topology to compare nodes, regardless of their distance in the network. This approach includes
some methods devoted to measure the topological regularity of networks~\cite{costa2009beyond,0295-5075-100-5-58002}. Similar approaches have also been used to provide a node-to-node mapping in general network analysis and in text analysis~\cite{Amancio20124406,doi:10.1142/S0129183108012285,PhysRevE.80.026103,0295-5075-100-5-58002}, as well as in pattern recognition~\cite{0295-5075-98-5-58001}.

\subsection{Global similarity}

The definitions presented so far
can only consider two nodes $i$ and $j$ as similar if they share a common neighborhood. However, in many real-world networks, nodes that do not share common neighbors can in fact play similar roles in network topology and therefore can be considered similar to each other~\cite{PhysRevE.73.026120}. Therefore, the definitions based solely on shared neighbors might be inappropriate to extract useful information about similarities in some networks.

Most of the measurements extending the concept of shared neighbors use shortest paths to quantify similarities. In the  measurement defined in~\cite{transitivo}, two nodes are considered similar to each other whenever they are connected by shortest paths involving low degree nodes. Mathematically, this measurement is given by~\cite{transitivo}
\begin{equation} \label{medidda}
    c_{ij}^{\ell} = \prod_{h} \frac{1}{k_h + k_{h+1} - 1},
\end{equation}
where the product is computed along the nodes belonging to the shortest path linking $i$ and $j$. Upon comparing systematically the accuracy of similarity measurements for link prediction in social networks, the authors showed in ~\cite{transitivo} that  Equation \ref{medidda} has advantages over other traditional measurements, without a significant loss in computational efficiency. The same measurement has been found to be useful to cluster nodes in graphs~\cite{5600463}.

A more complex conception of similarity considers that
two nodes are similar if their neighbors are similar. The basic idea consists in the definition
of the similarity index $c_{ij}^{\textrm{B}}$, whose value relies on the similarity between the neighbors of $i$ and $j$~\cite{ameasure,simrank,PhysRevE.73.026120}, given by
\begin{equation} \label{asdfw}
    c_{ij} = \alpha \sum_{kl} a_{ik} a_{jl} c_{kl} + \delta_{ij}
\end{equation}
or, in matrix terms, $\mathbf{C} = \alpha \mathbf{A} \mathbf{C} \mathbf{A} + \mathbf{I}$.
The iterative solution of this matrix equation, gives
\begin{equation}
	\mathbf{C} = \sum_{i=0}^\infty \alpha^i \mathbf{A}^{2i}.
\end{equation}
This means that only paths comprising an even number of nodes are used in the calculation of similarity. Evidently, there is no clear reason to ignore paths comprising an odd number of nodes. This problem is addressed by defining Equation \ref{asdfw} in a slightly different manner~\cite{newman2010networks}. According to this new definition, two nodes $i$ are $j$ are similar if $j$ has neighbors which are themselves similar to node $i$. Mathematically,
\begin{equation}
    c_{ij}^{\textrm{LHN}} = \alpha \sum_k a_{ik} c_{kj}^{\textrm{LHN}} + \delta_{ij}.
\end{equation}
The solution including paths of all lengths can be computed as
\begin{equation} \label{somainfinita}
    \mathbf{C}^{\textrm{LHN}} = \sum_{m=0}^\infty (\alpha \mathbf{A}) ^ m = (\mathbf{I} - \alpha \mathbf{A})^{-1}.
\end{equation}
This similarity index relies upon the choice of the parameter $\alpha$, which assigns the importance given for the longer paths. Whenever $\alpha \ll 1$, the similarity will depend mainly on the shortest paths. Applications of Equation \ref{somainfinita} include, for example, the computation of syntactical-semantical similarity measurements in texts modeled as complex networks~\cite{Amancio20124406}.

The similarity defined in Equation \ref{somainfinita} can be modified in several ways.
According to the Equation \ref{somainfinita}, high-degree nodes will tend to be more similar to other nodes than low degree nodes. As a consequence, the definition given by the Equation \ref{somainfinita} will present a bias towards high-degree nodes. To avoid such effect,
a straightforward modification in the formula could consider a normalization factor proportional to the degree $k_i$, that is,
\begin{equation} \label{anova}
    c_{ij}^{\textrm{LHN}} = \frac{\alpha}{k_i} \sum_k a_{ik} c_{kj}^{\textrm{LHN}} + \delta_{ij},
\end{equation}
or, equivalently, in matrix form
\begin{align}
    \mathbf{C}^{\textrm{LHN}} & =  \alpha \mathbf{D}^{-1} \mathbf{A} \mathbf{C}^{\textrm{LHN}} + \mathbf{I} \\ \nonumber
    & = (\mathbf{D} - \alpha \mathbf{A})^{-1} \mathbf{D}.
\end{align}
where $D_{ii} = k_i$ and $D_{ij} = 0$ for $i \neq j$.

Another modification in the similarity index established in Equation \ref{somainfinita} is to consider the expected number of paths in equivalent random networks~\cite{PhysRevE.73.026120}.
Expanding Equation~\ref{somainfinita} as a power series and normalizing the $n$-th term in the sum by the number of expected paths of length $n$ in a random network, the authors define the final form of the implicit equation for the similarity matrix $\mathbf{C}^{\textrm{LHN}}$ as~\cite{PhysRevE.73.026120}
\begin{equation}  \label{eq:S_matrix_form}
\mathbf{D}\mathbf{C}^{\textrm{LHN}}\mathbf{D} = \frac{\alpha}{\lambda_1}\mathbf{A}(\mathbf{C}^{\textrm{LHN}}\mathbf{S}\mathbf{D}) + \mathbf{I}.
\end{equation}
The authors in~\cite{PhysRevE.73.026120} show the potential of the new similarity measurement defined by Equation~\ref{eq:S_matrix_form}, applying it to the word network of the 1911 U.S. edition of \textit{Roget's Thesaurus}. The thesaurus consists of a hierarchical characterization of semantic linked words organized in different classes or levels of meaning. Thus, in the complex network mapping of the thesaurus, two words at the same level are considered to be connected if they have common words as entries in the previous level. In order to show the comparison between the similarity measurement defined by $\textbf{C}^{\textrm{LHN}}$ and the well-known cosine similarity defined in Equation~\ref{cosseno}, in Table~\ref{tab:roget} we reproduce the results obtained in~\cite{PhysRevE.73.026120} for the most similar words of ``alarm'', ``hell'', ``mean'' and ``water''. We can see that the measure defined by Leicht \textit{et al. } captures more general associations between words, whereas the cosine similarity is restricted to high values of similarity. This result can be explained by the fact that cosine similarity is proportional to the number of common neighbors, i.e., number of paths of length 2 between nodes. On the other hand, the definition in Equation \ref{eq:S_matrix_form} is based on paths with different lengths, encompassing the long range similarity between the nodes, justifying the better performance on quantifying the hierarchical organization of words classification.

\begin{table}
\tbl{The words most similar to ``alarm'', ``hell'', ``mean'' and
``water'' in the word network of the 1911 edition of \textit{Roget's
Thesaurus}, as quantified by the similarity defined in Equation \ref{eq:S_matrix_form} (setting $\alpha=0.98$) and by the more rudimentary cosine similarity (see Equation \ref{cosseno}).}
{\begin{tabular}{l|lr|lr}
\hline
  Word & \multicolumn{2}{c|}{Equation \ref{eq:S_matrix_form}} &
  \multicolumn{2}{c}{Cosine similarity} \\
\hline
        & warning &  32.014 & omen & 0.51640\\
  alarm & danger & 25.769 & threat & 0.47141\\
        & omen & 18.806 & prediction & 0.34816\\
\hline
        & heaven & 63.382 & pleasure & 0.40825\\
  hell  & pain & 28.927 & discontent & 0.28868\\
        & discontent & 7.034 & weariness & 0.26726\\
\hline
        & compromise & 20.027 & gravity & 0.23570\\
  mean  & generality & 19.811 & inferiority & 0.22222 \\
        & middle & 17.084 & littleness & 0.20101 \\
\hline
        & plunge & 33.593 & dryness & 0.44721\\
  water & air & 25.267 & wind & 0.31623\\
        & moisture & 25.267& ocean & 0.31623 \\
\hline
\end{tabular}}
\label{tab:roget}
\end{table}



\begin{figure}[!htbp]
\begin{center}
    \includegraphics[width=0.6\linewidth]{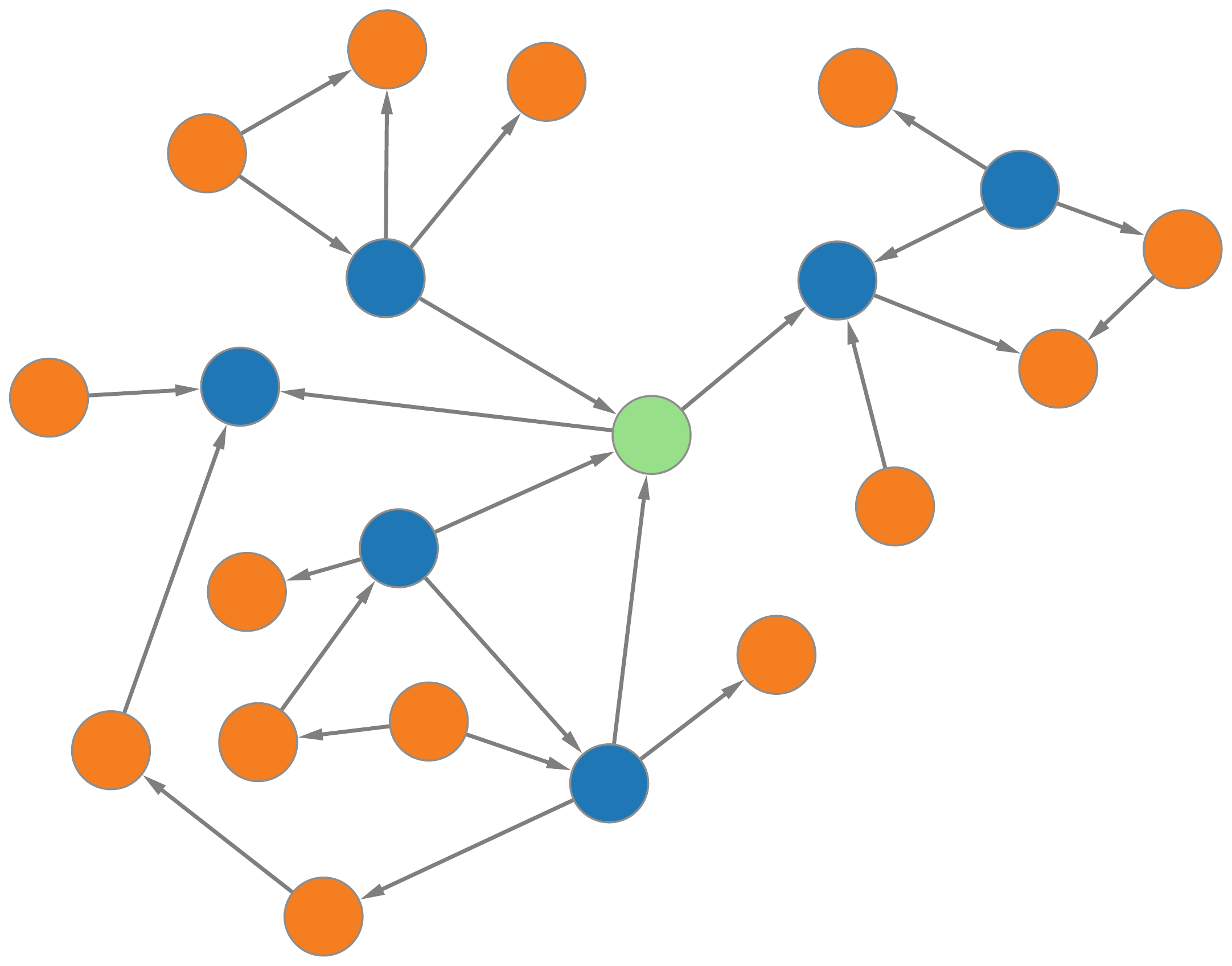}
        \caption{\label{f:vary_num_features}Example of growing process in the local citation network. The reference node is the green node. Nodes in blue and orange are the nodes belonging to the first and second layers, respectively.}
	\end{center}
\end{figure}

Other similarity measurements based on distances between nodes have been defined to tackle specific problems. In~\cite{Lu:2001}, the authors  suggest that the topological information should be employed along with semantic-based measurements to improve the characterization of directed acyclic networks, such as citation networks~\cite{newmancit}. The index proposed in \cite{Lu:2001} is based on the identification of both hubs and authorities in a subgraph around the two nodes $i$ and $j$ whose similarity is being estimated. More specifically, given two papers, the method constructs a local network for $i$ and $j$. The local networks are built from a growth process around a given node $i$. The first layer includes nodes that cites $i$ or nodes that appear in the reference list of $i$. In a similar manner, the second layer encompasses papers citing nodes in the first layer and nodes in the reference list of all papers in the first layer. Figure \ref{f:vary_num_features} illustrates the construction of a local network.
After the construction of the local subgraphs,  centrality indexes for each node in both local networks are computed. The similarity between the $i$ and $j$ is then estimated as the cosine of the vectors representing the centrality values of the neighborhood around $i$ and $j$. A normalization introduced before computing the cosine is useful to minimize the influence of hubs (e.g. surveys) that are similar to many other papers in the subgraph.




%

Similarity measurements have also been applied for the purpose of identifying topological communities~\cite{Pan20102849}. The basic idea behind the community detection techniques is that similar nodes tend to be clustered in the same community. In such cases, the similarity measurements are used to define a CSF transformation of the system. For example, the measurement proposed in~\cite{Pan20102849} is based on the concept of resource allocation in complex networks~\cite{refId0}. Given nodes $i$ and $j$, one assumes that $i$ is able to send resources to $j$ through its neighbors. Considering that each node is able to handle a fixed amount of resources, the similarity between $i$ and $j$ is expressed as
\begin{equation} \label{enough}
    c_{ij} = \sum_{z \in | \Gamma_i \cap \Gamma_j |} {k^{-1}_z}.
\end{equation}
If  $z \in | \Gamma_i \cap \Gamma_j |$ has only two connections, i.e. $k_z = 2$, then all the information leaving $i$ reaches $j$, and the contribution of $z$ for $c_{ij}$ is maximum. Conversely, if $z$ has many neighbors, then only a small fraction of the information leaving $i$ reaches $j$.
A problem arising from the definition in Equation \ref{enough} is that the similarity does not take into account a possible direct link between $i$ and $j$. To overcome this, a simple modification was proposed in~\cite{Pan20102849}, defining the similarity as

\begin{equation} \label{modified}
    c_{ij} =
    \begin{cases}
    k^{-1}_z & \textrm{if $i$ and $j$ are neighbors,} \\
    0 & \textrm{otherwise.}
    \end{cases}
\end{equation}



Some similarity can also be defined in terms of random walks~\cite{doi:10.1142/S0217979215500952}. 
Given an initial node $i$, a neighbor $j$ is chosen randomly as the next node to be visited. The transition probability between nodes $i$ and $j$ is given by
\begin{equation} \label{normalization}
    p_{ij} = w_{ij} / \sum_{k \in \Gamma(i)} w_{ik},
\end{equation}
where $w_{ij}$ is the weight of the edge connecting nodes $i$ and $j$.
Let $p_{ij}^{(m)}$ be the probability that the walker, departing from node $i$, reaches node $j$ in $m$ steps. The transition probabilities to each node can be represented as the vector $\mathbf{p}_i^{(m)}$, where the $j$-th element of $\mathbf{p}_i^{(m)}$ is given by $p_{ij}^{(m)}$. An important property concerning the stationary distribution of $p_{ij}^{(m)}$ states that, in an undirected network
\begin{equation}
    \lim_{m \to \infty} p_{ij}^{(m)} = \frac{k_j}{2E},  \ \ { \forall i \in V}.
\end{equation}
This means that, if $m$ takes values much higher than the mixing time~\cite{norris1998markov}, then $\mathbf{p}_i^{(m)}$ will not quantify a similarity but a structural property of the target node.


An application of random walks for the computation of topological node similarity is provided in~\cite{Harel:2001}.
In their study, the authors define a similarity index aiming at recognizing patterns in spatial data.  Given two nodes $i$ and $j$, they compare the values of $\mathbf{p}_i^{(m)}$ and $\mathbf{p}_j^{(m)}$ using standard similarity/dissimilarity measurements. An example of distance defined by the authors is
\begin{equation} \label{computat}
    d(i,j) = \exp\Big{(}2m - \|\mathbf{p}_i^{(m)} - \mathbf{p}_j^{(m)}\|_L\Big{)} - 1,
\end{equation}
where $\| \mathbf{x} - \mathbf{y} \|_L = \sum |x_i - y_i|$. In the computation of Equation \ref{computat}, the authors do not consider a specific length $m$. Instead, they use several values of $m$ because, in a bipartite subgraph, random walks starting at $i$ and $j$ might not visit the same nodes for a specific value of $m$. 


A measure similar to the one defined in Equation \ref{computat} is defined in~\cite{Pons04computingcommunities} for the purpose of community detection in complex networks. The authors define their measurement taking into account the following remarks: (i) if two nodes $i$ and $j$ belong to the same topological community, then $p_{ij}$ will take high values. The counterpart affirmation does not hold, i.e. high values of $p_{ij}$ do not imply that $i$ and $j$ share the same community. For example, high degree nodes tend to cause $p_{ij}$ to take high values even if they are placed at distinct communities. (ii) two nodes belonging to the same community display similar topological properties. As a consequence, if $i$ and $j$ belong to the same community, then $p_{ik}^{(t)} \simeq p_{jk}^{(t)}$. (iii) If $j$ is highly connected, then the probability $p_{ij}^{(t)}$ will probably take high values, because the walker will be able to access $j$ through many paths. Based on these three remarks, the similarity between nodes $i$ and $j$ was defined as~\cite{Pons04computingcommunities}
%
%
%
%
\begin{equation} \label{sm1}
    r_{ij}^{(t)} = \sqrt{\sum_{l} \frac{ ( p_{il}^{(t)} - p_{jl}^{(t)} )^2 }{k_l} } = \| \mathcal{D}^{-\frac{1}{2}} \mathcal{P}_{i}^{(t)} - \mathcal{D}^{-\frac{1}{2}} \mathcal{P}_{j}^{(t)} \|,
\end{equation}
where $\mathcal{D} = \{\delta_{ij}\}$ is a diagonal matrix. The element $\delta_{ij}$ is given by
\begin{equation*}
\delta_{ij} = \left\{
\begin{array}{rl}
k_i, & \textrm{if } i = j,\\
0, & \textrm{otherwise.}\\
\end{array} \right.
\end{equation*}
The measurement defined in Equation \ref{sm1} can also be written in terms of the spectrum of matrix $P$~\cite{Pons04computingcommunities} as
\begin{equation}
    r_{ij}^2 = \sum_{\alpha=2}^N \lambda_\alpha^{2t} (v_\alpha(i) - v_\alpha(j))^2,
\end{equation}
where  $\lambda_\alpha$ and $v_\alpha$ ($1 \leq \alpha \leq N$) are given by
\begin{equation}
    P v_\alpha = \lambda_\alpha v_\alpha.
\end{equation}
Matrix $P$ is called the transition matrix of the random walk. The definition in Equation \ref{sm1} can also be extended to consider the distance between communities. To do so, the probability of the walker going from community $C_k$ to node $j$ is defined as
\begin{equation} \label{a432}
   p_{C_k}^{(t)} = \frac{1}{|C_k|} \sum_{i \in C_k} p_{ij}^{(t)}.
\end{equation}
Using Equation \ref{a432}, the distance between two groups of nodes (communities) is given by
\begin{equation}
    r_{C_kC_l}^{(t)} = \sqrt{ \sum_{i=1}^N \frac{ (p_{C_ki}^{(t)} - p_{C_li}^{(t)})^2 }{d(i)} } = \|| D^{-\frac{1}{2}} P_{C_k}^{(t)} - D^{-\frac{1}{2}} P_{C_l}^{(t)} \|.
\end{equation}
%



Random walks have also been applied to categorize words through navigation in semantic networks. In the research carried out by Holthoefer et. al.~\cite{categorizing}, the authors perform random walks in a free association network~\cite{categorizing} according to Equation~\ref{normalization}.
More specifically, random walks of variable length are considered in the matrix storing frequencies of access in terms of probability values, that is,
\begin{equation}
    \mathbf{T} = \lim_{ S\rightarrow\infty } \sum_{i=1}^S \mathbf{P}^i = (I - P)^{-1}
\end{equation}
In other words, each node is represented by the frequency of access in random walks with length ranging from $1$ to $S$. The similarity between two nodes is then computed as the cosine of the vectors representing the nodes in the new space, i.e.
\begin{equation} \label{novvva}
    \mathbf{C} = \mathbf{T} \mathbf{T} ^ {T}.
\end{equation}
The authors emphasize that the measurement defined in Equation \ref{novvva} differs from other long-range connectivity-based measurements because it does not rely on the \emph{number} of paths, but on the navigation properties of  semantical networks. Among the capabilities of this measure, it has been found that it is able to detect words semantically similar. In addition, this measurement has proven useful to map free associations networks with heterogeneous links into semantic networks comprising links of the same nature~\cite{categorizing}. 



%

Some node similarity measurements rely on the computation of the rank-k matrix $\mathbf{M}_k$, which is an approximation of the original adjacency matrix $\mathbf{A}$~\cite{Liben-Nowell:2003}. An efficient and widely employed method to compute $M_k$ is the singular value decomposition (SVD)~\cite{svdref}, which is at the core of the latent semantic analysis~\cite{LSA-ARIS} and pseudo-inverse computations~\cite{penrose}.
%
%
Matrix $\mathbf{M}_k$ might then be employed to compute the similarity, for example, in Equations \ref{cosseno} and \ref{somainfinita}. Methods based on low-rank approximations have proven useful for the analysis of real networks because they map the adjacency matrix into a more simplified representation able to reduce the amount of noise present in the original matrix representation.




\section{Topology embedding}

The transformation CF, shown in Figure~\ref{f:triangle}, involves using the adjacency matrix to directly define node features. The main use of such procedure is to visualize networks, therefore here we focus on different methods aiming at providing a clear and intuitive visualization of the system topology. Nevertheless, the CF transformation is also important in some community detection methods. The Fiedler method~\cite{fiedler1973algebraic,pothen1990partitioning} transforms the Laplacian matrix of the graph into a vector (the second smallest eigenvector for connected graphs) that contains a good guess of the optimal partition of the graph. The so-called \emph{modularity matrix} can also be used to define successive bisections of the network~\cite{newman2006modularity,newman2006finding}, and is a widely used method for detecting communities in small networks.

\subsection{Principal component analysis}
\label{subsec:pca}
Principal component analysis (usually called PCA) is one of the oldest forms of multivariate analysis~\cite{jolliffe2002principal}. One of the earliest mentions to this technique is by Pearson in 1901 \cite{Pearson:1901}, while more formal definitions were given by Hotelling in 1933~\cite{Hotelling:1933}. The main idea of PCA is to reduce the dimension of a given dataset by eliminating redundancy between its variables. This is done by projecting the original $m$ variables into a new basis of dimension $m'\leq m$, while preserving as far as possible the variance of the data. In more formal terms, given a $m$-dimensional vector $\vec{x}$ containing the characteristics of an object, we seek a vector $\vec{\eta}_1$ so that the variance of the projection $var[\vec{\eta}_1 \vec{x}]$ is maximal. Then, we seek a new vector $\vec{\eta}_2$, uncorrelated with $\vec{\eta}_1$ and where $var[\vec{\eta}_2 \vec{x}]$ is maximal. This process is repeated until $m'$ vectors are obtained. This means that the features of the original objects are transformed into a similarity measure, which in turn defines new features for the objects. Therefore, the procedure follows the FSF path defined in Figure~\ref{f:all_paths}.

Pearson, and independently Hotteling, found that the basis of maximal variance ${\vec{\eta}_1,\vec{\eta}_2,...,\vec{\eta}_m}$ is defined by the equation~\cite{jolliffe2002principal}

\begin{equation}
	\Sigma \vec{\eta}=\lambda \vec{\eta} \label{eq:autoPCA}
\end{equation}
where $\Sigma$ is the covariance matrix obtained over distinct realizations of $\vec{x}$. These realizations are usually measurements taken over distinct objects. Equation \ref{eq:autoPCA} is a typical eigenvalue equation defining eigenvalues $\lambda_1,\lambda_2,\dots,\lambda_m$ and eigenvectors $\vec{\eta}_1,\vec{\eta}_2,\dots,\vec{\eta}_m$. The eigenvalues $\lambda_i$ are known to represent $var[\vec{\eta}_i \vec{x}]$. Therefore, ordering the eigenvectors $\vec{\eta}_i$ according to the decreasing order of the respective eigenvalues, i.e., $\lambda_1 > \lambda_2 > ... > \lambda_m$ allows us to write~\cite{jolliffe2002principal}

\begin{equation}
   \vec{z}=E' \vec{x}, \label{eq:proj_PCA}
\end{equation}
where the $i$-th column of matrix $E$ is the eigenvector $\vec{\eta}_i$, and $E'$ means the matrix transpose of $E$.


PCA can be used to visualize and analyze networks. In~\cite{costa2009seeking} the technique was used to develop the concept of \emph{regularity} of complex networks. The idea is that nodes in a so-called \emph{complex} network usually display a large range of values for a given set of topological measurements. This is a striking contrast to the simplicity of traditional topological spaces (e.g., orthogonal and triangular lattice), where the neighborhood of all nodes have the same topological characteristics. In order to measure the regularity of a network, the authors used four topological measurements to characterize the nodes, and applied the PCA to project the network into a 2D space. This can provide a visual understanding of the variability of the measurements for different network nodes, as well as unveil the presence of clusters in the network (i.e., nodes having similar characteristics). The procedure defines a CSF path, since its main objective is to create a visual representation of the relationships between nodes. Furthermore, the authors defined a more quantitative evaluation of regularity by calculating the probability density distribution of the points in the PCA space, which was done using the non-parametric Parzen window method~\cite{duda2012pattern}. One of the main insights of the analysis is that regularity does not necessarily implies a narrow degree distribution. The authors also found that the internet and protein-protein interaction networks seem to be getting simpler over time. 

In a related work~\cite{da2010pattern}, the authors generalized the PCA analysis presented in~\cite{costa2009seeking}, and used the PCA technique to characterize networks in three distinct levels of detail, defined by the whole network, by the communities in the network and by node characteristics. A similar technique has been used to detect singular motifs in networks~\cite{costa2009beyond}, which are unique structures in the network having a clear functional role in the system.

As explained above, by means of a linear combination between the original measurements the PCA can provide a new set of measurements to describe nodes in the network in a more concise form. This property was used in~\cite{rodrigues2010generalized} to define a measurement of generalized connectivity of a node. The authors defined a three dimensional matrix $M$ containing at row $i$, column $j$ and depth $k$ the number of paths between nodes $i$ and $j$ with length $k$. Each node $i$ is characterized by the $i$-th row of matrix $M$ for a given plane $k$ of the matrix. Therefore, the respective network can be projected into a 2D or 3D space for different values of $k$. It was shown that the two communities in the Zachary Karate club network~\cite{zachary1977information} can be exactly recovered by projecting the network for $k=2$. Regarding community detection, the PCA has also been used to detect functional brain modules for fMRI time series~\cite{leonardi2013principal}.
\subsection{Network visualization}
\label{sec:networkVisualization}

The visualization of a network is commonly accomplished by constructing a diagram of symbols and lines which are placed over a $2$ or $3$-dimensional metric space representing respectively its nodes and edges. While the process of choosing symbols and the overall aesthetics for the visualization is a general problem in the computational visualization field~\cite{tollis1999graph}, the problem of assigning topologically meaningful positions to nodes are traditionally studied in the fields of graph theory or network science, being commonly referred to as \emph{graph drawing}.

In general, a graph drawing on a $d$-dimensional metric space for a network $\mathcal{G}(\mathcal{N},\mathcal{E})$ can be defined as a \emph{layout map} $M:\mathcal{N}\rightarrow\mathbb{R}^d$. A formal embedding (also known as a \emph{strong embedding}) of a graph~\cite{cohen1994three} is a map restricted by the fact that no crossing among edges occurs. Only a certain class of graphs, known as planar graphs, can be strongly embedded into $\mathbb{R}^2$, on the other hand, any graph is strongly embeddable into $R^3$~\cite{trudeau2013introduction}.

The issue of having crossing edges in a graph drawing is relevant to many problems of graph theory~\cite{gansner1993technique,eades1994edge} and certain particular problems such as in the case of printed circuit board design~\cite{quinn1979forced}. However, since the vast majority of complex networks representing real systems are not perfectly planar, the constraint of non-crossing links is usually not enforced. Therefore, in network science, an embedding (or \emph{weak embedding}) refers to any map $M:\mathcal{N}\rightarrow\mathbb{R}^d$ of a network $\mathcal{G}$.

The process of choosing an approach to embed a network through a meaningful map depends on the purpose of the graph drawing. Conventionally, this choice is made by selecting the characteristics which are preferable to be preserved or emphasized in the drawing. Therefore, many graph drawing techniques have been defined in the literature for a great variety of goals. Among the most popular methods are the spectral~\cite{hall1970r,harel2002graph, koren2005drawing}, force-directed~\cite{eades1984heuristics,kamada1989algorithm,Fruchterman1991Gr}, distance minimization~\cite{kruskal1964multidimensional,paulovich2008least,viana2011fast}, hierarchical~\cite{tamassia2014handbook} and circular~\cite{baur2005crossing} layouts. Distinctly from the others, the last two are representatives of a class of techniques with strict constraints under the regions permitted for occupation of nodes. For instance, in circular layout, nodes can only occupy positions in a circle. In contrast, force-directed methods usually permits nodes to occupy any position in the target metric space.

Hierarchical layout methods can be applied to networks having a well-defined hierarchical organization, such as trees and directed acyclic graphs. The main objective of such kind of layout is to place nodes of same hierarchical level within the same specified regions, such as lines, planes or concentric circles; and minimizing the number of crossings among edges of distinct levels~\cite{tamassia2014handbook}. For networks with no explicit hierarchical organization, the nodes hierarchy can be inferred by suitable techniques. For instance, the degree of each node can be used, so that nodes having the same degree lie in the same hierarchical level. Other examples include the use of $k$-core decomposition~\cite{alvarez2005large} to rank the hierarchy or by considering the structure of the minimum spanning tree of a network.

In a circular layout, nodes are regularly placed in a circle. Usually, the order of the placement (i.e. the angle) is used to emphasize the patterns of a particular scalar characteristic~\cite{baur2005crossing}. Node degree, for instance, can be used as a ranking parameter to reveal the patterns of connections among nodes with similar degree. Similarly, the community structure of the network or the category of nodes can also be used to organize them around the circle, resulting in an overview of how distinct groups are connected among themselves.

The aforementioned layout methods are useful for generating particular visualizations of a network, however they hold little information about its topological structure. Methods for graph drawing that preserves the topological structure of networks are referred to as \emph{topology embedding} and can be categorized into two main groups: partial embedding methods~\cite{kondor2013measuring} and complete embedding methods~\cite{kruskal1964multidimensional,paulovich2008least}. A partial embedding of a network preserves the adjacency among connected pairs of nodes by bringing then close together. At the same time, it places unconnected pairs of nodes as far as possible. In contrast, other techniques based in the complete embedding of the network, such as multidimensional scaling (see Section~\ref{sec:mult_scaling_graphs}) or the classic Kamada-Kawai algorithm~\cite{kamada1989algorithm}, utilizes the complete set of shortest paths among nodes in a network. Furthermore, while a partial embedding is usually executed directly over the target space (usually a 2D or 3D metric space), a complete embedding demands spaces of much higher dimension. Consequently, complete embeddings require an extra step to account for a projection or a dimensionality reduction. Another way to define a partial embedding is through the use of spectral properties of graphs. For instance, a suitable layout map can be directly obtained from the eigenvectors of the Laplacian matrix of graphs~\cite{hall1970r,koren2005drawing} (see Section~\ref{sec:spectralEmbedding}). 

A perfect partial embedding employs a map where the distances among connected pairs of nodes are all equal to a fixed unit. Furthermore the distances among non connected nodes are maximized. Because the latter condition is impossible to attain in the vast majority of complex networks mapped to a 2D or 3D space, the first condition is often relaxed. Nevertheless, the distances among connected nodes are optimized to be as close as possible to the expected fixed unit. Force-directed layouts are among the most prominent classes of embedding methods used to construct partial or complete embeddings of networks. These methods are founded in the idea that graphs can be modeled as real physical systems, and that the dynamics of such systems can be used to determine the positions of nodes corresponding to an embedding.

A complete embedding takes into consideration the entire topology of a network, this means that much more information about the its structure is mapped to the target space. Usually, this is accomplished by first mapping each possible relationship among nodes by a similarity or distance measurement. From the resulting relationship map the positions of nodes on the target space is obtained by employing a metric embedding technique. For instance, the lengths of the shortest paths between nodes in a network can be used as an input distance matrix to a metric embedding algorithm, such as multidimensional scaling. Sections~\ref{sec:local_similarity} and~\ref{sec:mult_scaling_graphs} explore in detail, respectively, the methodologies based on constructing similarity relationships from networks and methods to generate positions from distance or similarity relationships.

\begin{figure*}[!tpb]
  \begin{center}
\begin{center}
\subfigure[Circular layout]{\label{fig:circularLayout} \includegraphics[width=6.0cm]{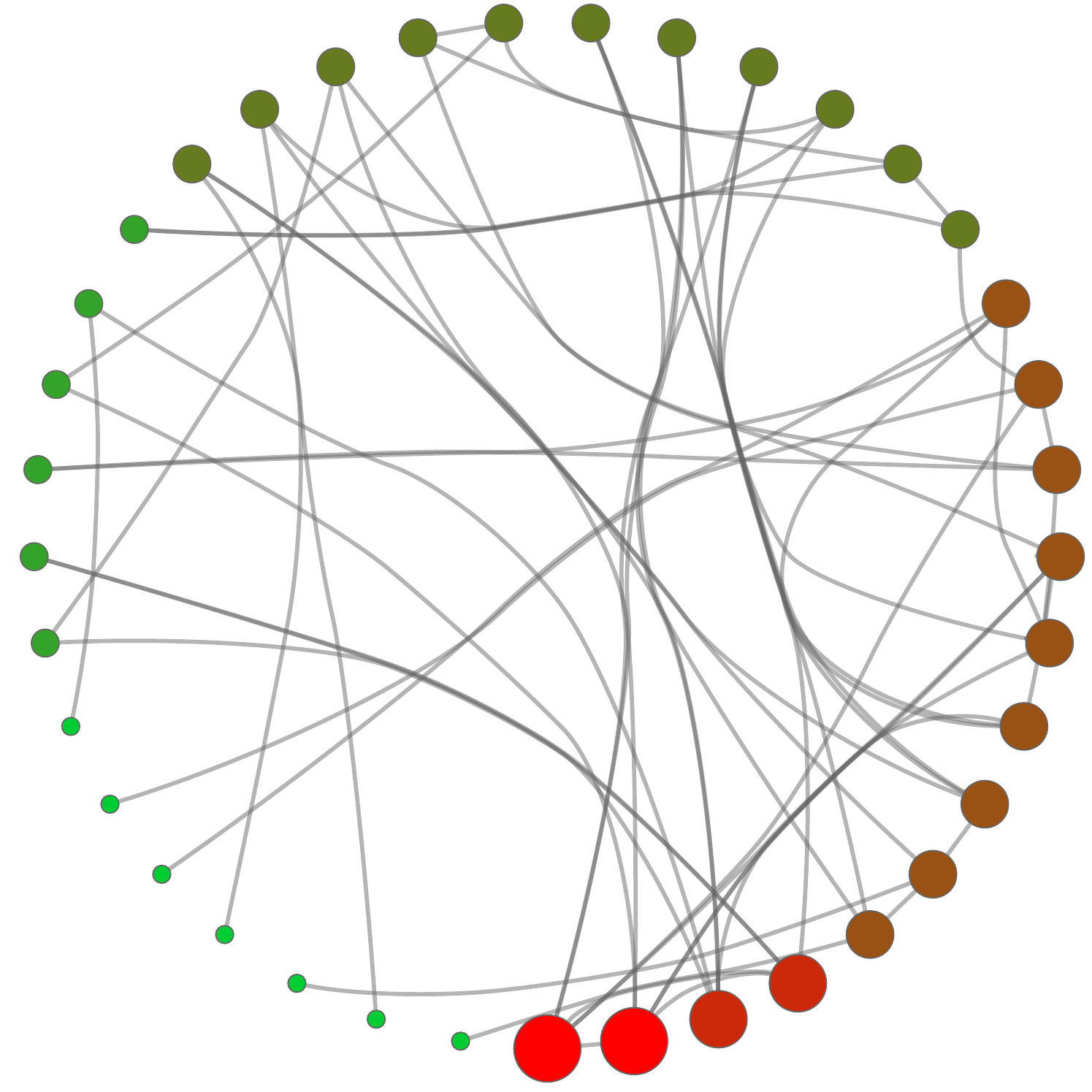} }~
\subfigure[Hierarchical layout]{\label{fig:hierarchicalLayout} \raisebox{0.60cm}{\includegraphics[width=6.0cm]{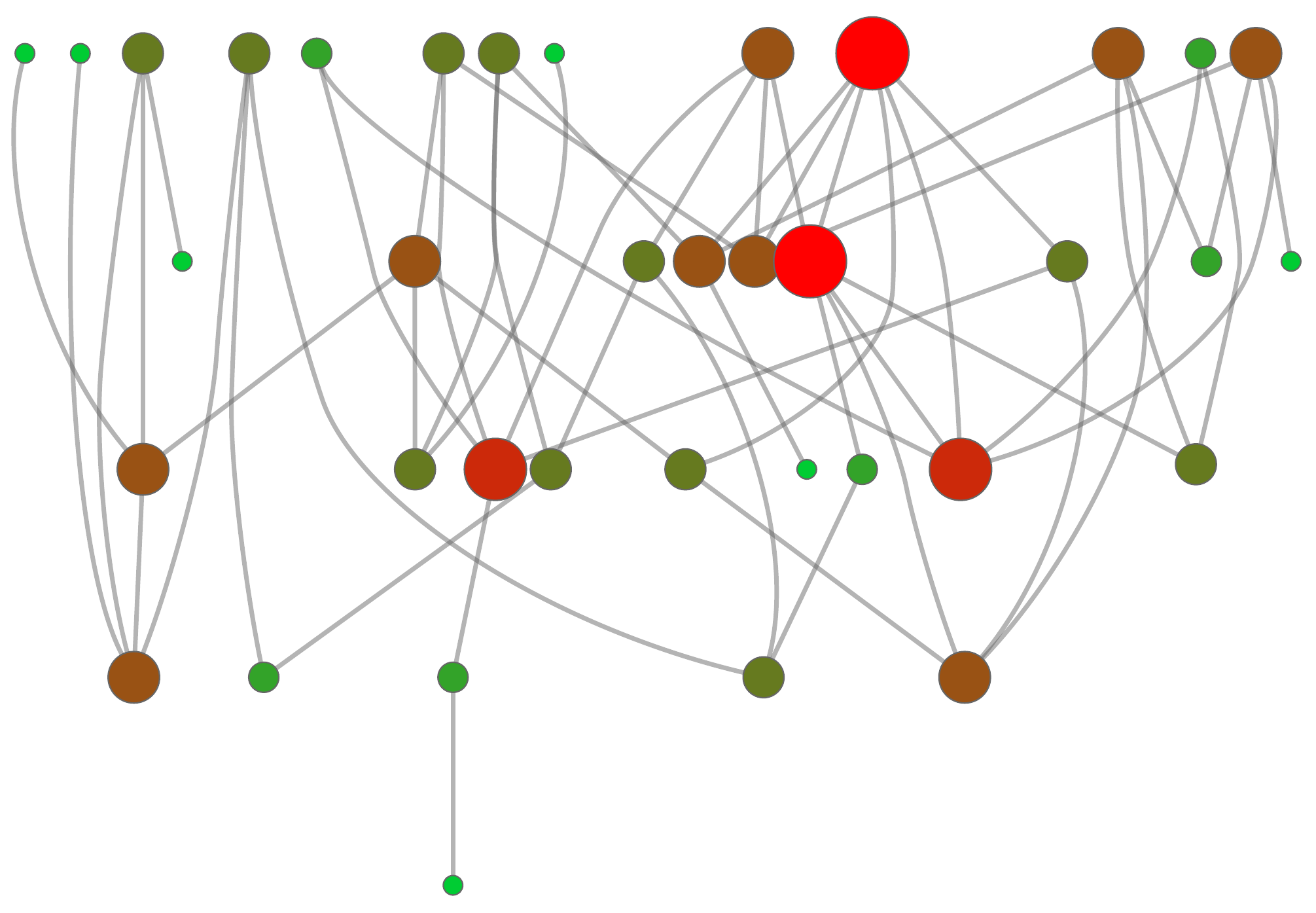}} }\\
\subfigure[Spectral layout]{\label{fig:spectralLayout} \raisebox{0.30cm}{\includegraphics[width=6.0cm]{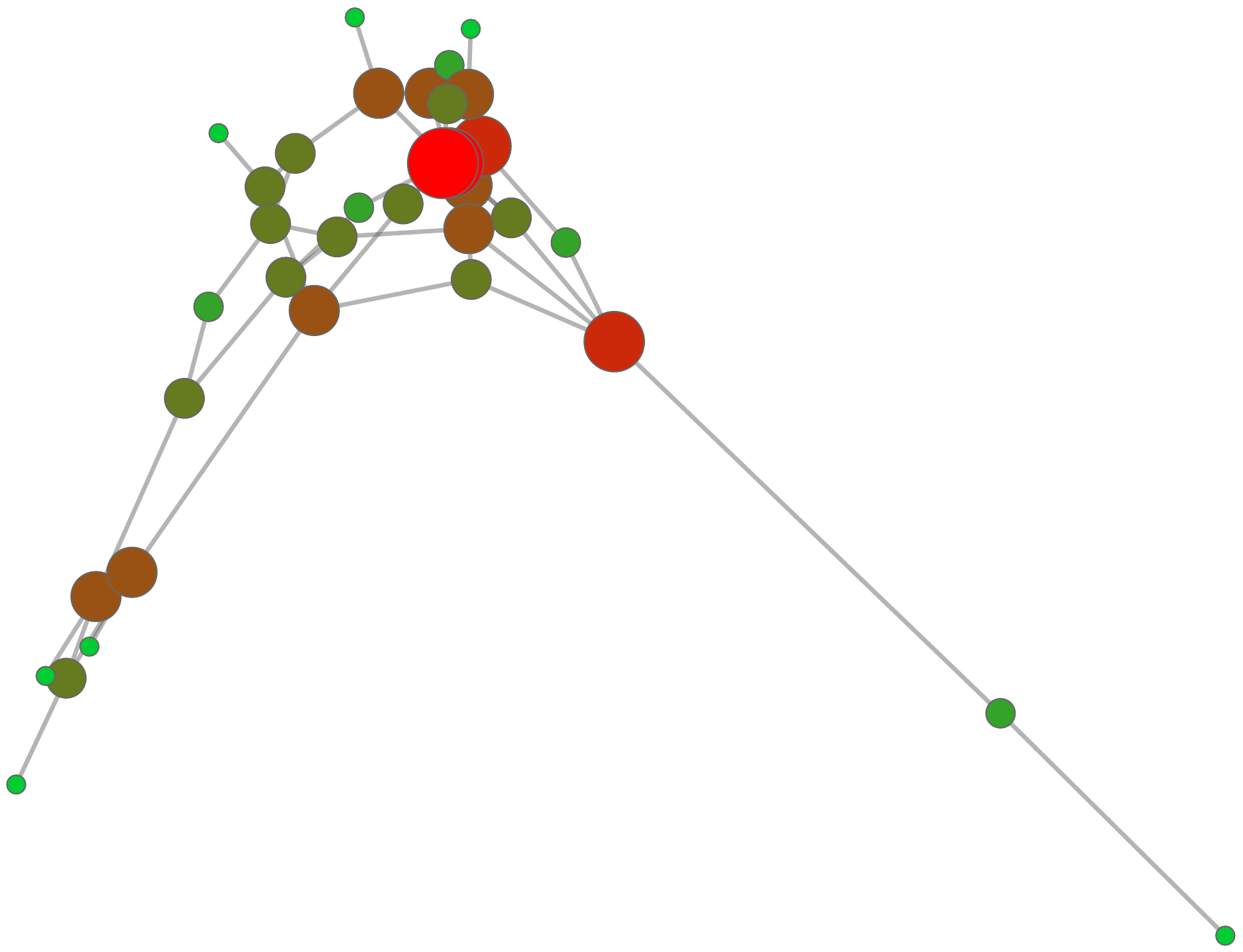}} }~
\subfigure[Force-directed layout]{\label{fig:topologicalLayout} \includegraphics[width=6.0cm]{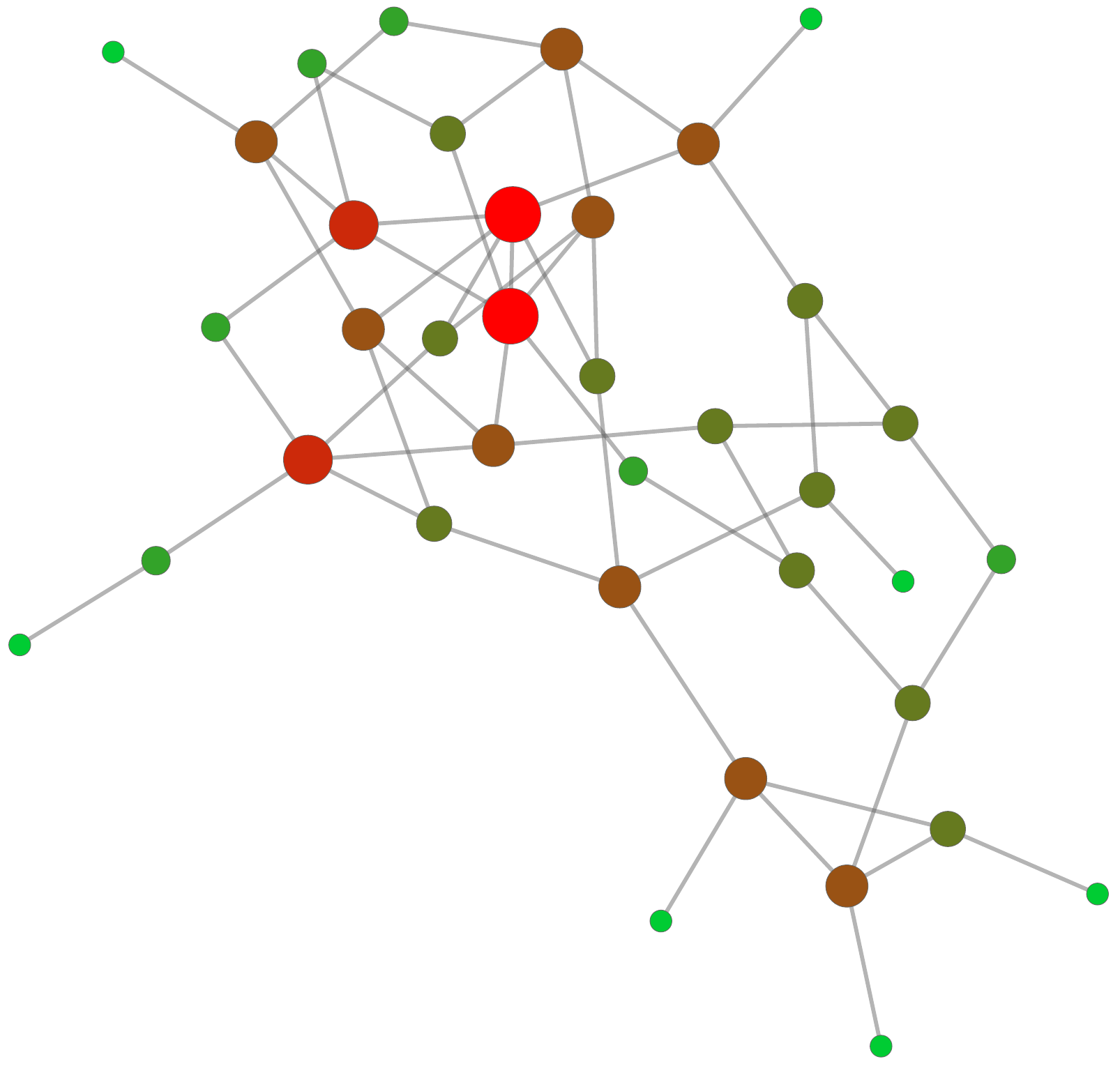} }
\end{center}
  \caption{Examples of distinct layouts obtained for the same network. Node colors and size indicate the degree of nodes.}
  ~\label{fig:layouts}
  \end{center}
\end{figure*}

Figure~\ref{fig:layouts} illustrates a few distinct layouts obtained for the same network. Note that, for the four method shown, only spectral and force-directed methods preserve the adjacency of nodes. The spectral map nevertheless present a less regular distribution of distances when compared to the force-direct layout, this is a known common artifact of spectral graph drawings~\cite{koren2003spectral,hachul2005experimental}.

\subsubsection{Force-directed topology embedding}
\label{sec:forceDirectedtopologyEmbedding}

The idea of using physical forces to layout a graph was initially proposed by Quinn and Breuer~\cite{quinn1979forced}. The authors considered graphs as dynamical physical systems consisting of objects and springs. In this system, nodes become moving objects and edges are transformed into springs, so that connected nodes interact according to Hooke's law. In addition, a repulsive force among all pairs of nodes is used to account for the maximization of the distances between unconnected nodes. Eades improved the method by utilizing non-realistic physical forces which are distinct from the previously employed Hooke's law~\cite{eades1984heuristics}. This deviation from the physical system was an important step toward the optimization of the method, owing to the fact that the calculation of square roots and general floating point arithmetics were computationally costly at that time. Both Eades' and Quinn-Breuer methods result in partial embeddings of a network.

On the other hand, Kamada and Kawai took the opposite direction and proposed a new method of force-directed complete embedding, known as \emph{Kamada-Kawai layout} (\emph{KK})~\cite{kamada1989algorithm}. In the KK method all topological distances among nodes are considered. Therefore, all pairs of nodes, being them connected or not, can be considered as having a spring between them. The relaxed distance $d_{ij}$ for each spring depends on a dissimilarity measurement taken over its respective pair of nodes $(i,j)$. The forces associated to each spring follows Hooke's law, thus the total energy $E(\mathcal{G})$ of a graph $\mathcal{G}(\mathcal{N},\mathcal{E})$ for the KK configuration is defined as
\begin{equation}
\label{eq:KamadaKawaiEnergy}
	E(\mathcal{G}) = {\sum\limits_{i\,\in\,\mathcal{N}}\sum\limits_{j\,\in\,\mathcal{N}} k ( |\mathbf{r}_i - \mathbf{r}_j| - d_{ij})^2},
\end{equation}
where $k$ is the spring constant and $\mathbf{r}_i$ the position of node $i$. The relaxed distance $d_{ij}$ is usually taken as the length of the shortest path between nodes $i$ and $j$, but other dissimilarity metrics can also be employed. 

The KK algorithm starts by randomly placing nodes over the target space. Next, the energy $E(\mathcal{G})$ is minimized so that the distances $|\mathbf{r}_i - \mathbf{r}_j|$ approach the preferred distances $d_{ij}$. Instead of solving the equivalent dynamical equations of the system, the method uses a greedy optimization approach. For every iteration, a node is chosen and moved to a new location corresponding to its minimum energy contribution in that configuration. The process is repeated until the energy reaches a minimum or a specified threshold. Methods based on the KK algorithm lie in the class of repeating CSF procedures. 

Because the KK method have many constraints and uses a greedy approach to optimize the energy, when targeting low dimension spaces, the resulting drawing is often a poor local minimum of the energy function. In some situations, such as for very large networks, this effect undermines the representation of the overall topology and even of the adjacency of nodes.

\begin{figure*}[!tpb]
  \begin{center}
  \includegraphics[width=0.95\linewidth]{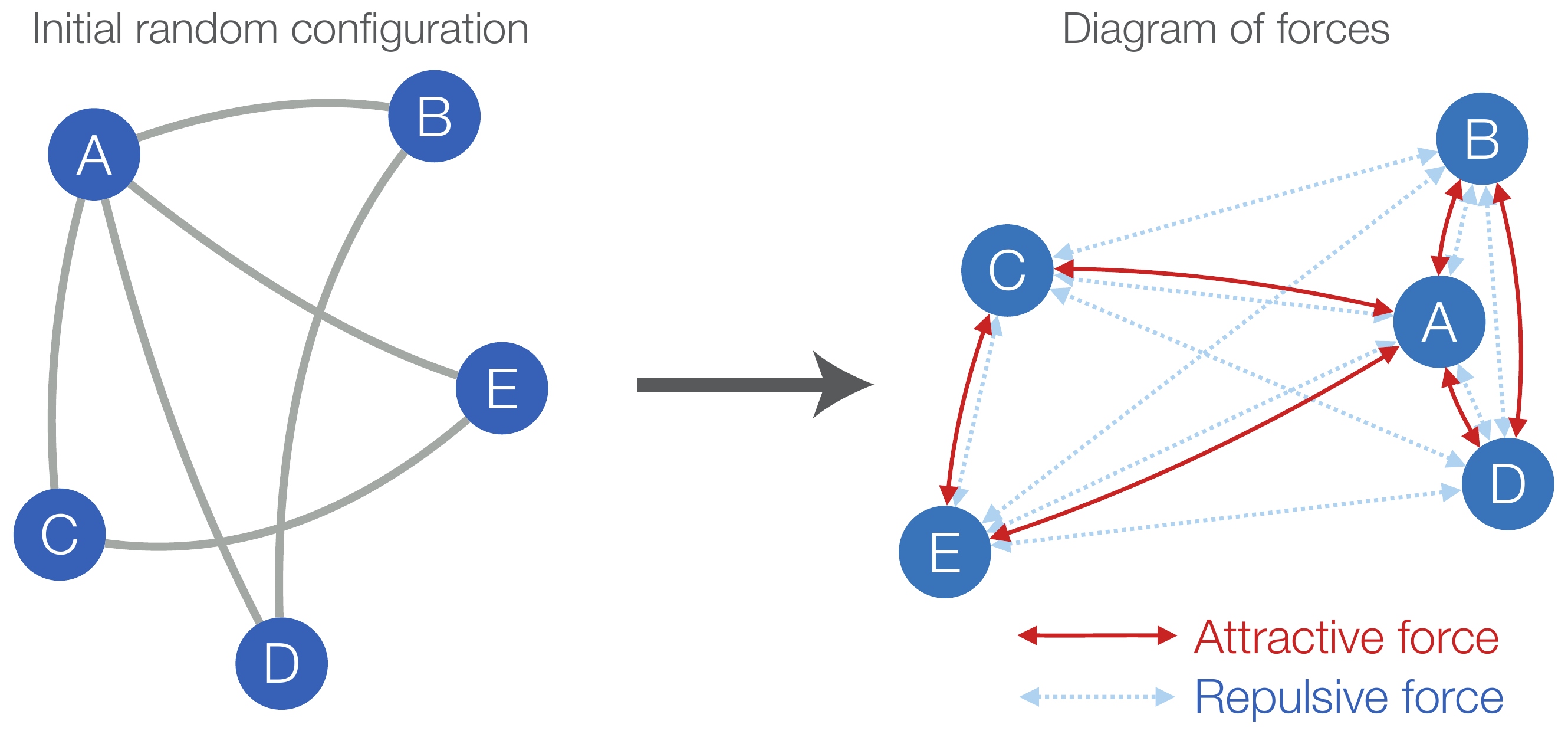} 
  \caption{Diagram of forces for a graph on a FR embedding.}
  ~\label{fig:fig_FR_example}
  \end{center}
\end{figure*}

Fruchterman and Reingold overcome the problem of poor local minima by proposing a new partial embedding method, known as the \emph{Fruchterman-Reingold layout} (\emph{FR})~\cite{Fruchterman1991Gr}. In the FR method, each pair of nodes $(i,j)$ is allowed to interact via two simple forces: a repulsive force $\mathbf{f}_{ij}^{(r)}$ and an attractive force which is added only for connected nodes $\mathbf{f}_{ij}^{(a)}$. These forces, originally not based on any real physical system, are given by
\begin{eqnarray*}
\label{eq:FRforces}
\mathbf{f}_{ij}^{(r)} &=& - (d^*)^2\frac{1}{|\mathbf{r}_i - \mathbf{r}_j|} \hat{r}_{ij} \\
\mathbf{f}_{ij}^{(a)} &=& {d^*} [\mathbf{r}_i - \mathbf{r}_j]^2 \hat{r}_{ij}
\end{eqnarray*}
where $d^*$ accounts for the optimal distance among nodes and $\hat{r}_{ij} = \frac{\mathbf{\mathbf{r}_i - \mathbf{r}_j}}{|\mathbf{r}_i - \mathbf{r}_j|} $. Therefore, the total force $\mathbf{F}_i^{(\text{total})}$ acting on each node $i \in \mathcal{N}$ is
\begin{eqnarray*}
\label{eq:FRforcesTotal}
\mathbf{F}_{i}^{(\text{total})} &=& \sum\limits_{j\,\in\,\mathcal{N}}  \mathbf{f}_{ij}^{(r)} +\sum\limits_{(i,j)\,\in\,\mathcal{E}} \mathbf{f}_{ij}^{(a)}.
\end{eqnarray*}
Figure~\ref{fig:fig_FR_example} illustrates the forces involved in a FR embedding for a simple network.

The dynamics however are undertaken differently from a realistic physical system. Instead of solving the equivalent differential equations of motion, at each iteration all nodes are allowed to move over the target space in the direction given by $\mathbf{F}_{i}^{(\text{total})}$. Nevertheless, the maximum displacement of a node is limited by a maximum value $\delta(T)$, which is determined by a ``temperature'' $T$ that decreases over time. The temperature is introduced so that the algorithm starts with a coarse-grained dynamics (i.e. larger $\delta(T)$) which gradually becomes more fine-grained (i.e. smaller $\delta(T)$). This procedure bears similarities with the simulated annealing method~\cite{szu1987fast}. However, instead of allowing the system to attain states of higher energy with small probability, the FR optimization method always go straight downhill along the energy space. It is not surprising that simulated annealing was also employed by other researchers to achieve better local minima for force-directed methods~\cite{davidson1996drawing} despite its higher computational costs.

In most cases, the FR algorithm results in graphically pleasing visualizations of networks. Common real network patterns such as community structure and geographical constraints emerge naturally from the use of this technique. The main drawback of the method lies in its computational cost. In particular, the complexity for calculating the repulsion among all nodes for an iteration is $O(|\mathcal{N}|^2)$. Thus, the method may not be suitable for large networks. Consequently, over the years several optimizations to the FR method have been proposed~\cite{hu2006Efficient,hachul2004drawing,hachul2005experimental}. Most of these optimizations relies in subdividing the target space in regions and calculating the repulsive forces only among pairs of nodes residing in the same region. Another approach to optimize the algorithm is to progressively simplify the network as a group hierarchy of nodes and calculate the forces in a multi-level approach~\cite{hachul2004drawing}. In general, a combination of these solutions is necessary to achieve satisfactory force-directed embeddings for large networks with more than hundreds of thousand of nodes~\cite{hu2006Efficient}. 

Because of the versatility of force-directed layout techniques, the FR algorithm can be easily adapted to emphasize other properties aside from the topological structure. For instance, the method can be modified so that the repulsive force among pairs of nodes is proportional to their degrees ~\cite{jacomy2014forceatlas2}, resulting in an embedding where hubs are more scattered over the space.

In the same fashion as the KK method and other force-directed techniques, the FR algorithm can be considered, internally, as an initial CF procedure followed by a repeating FCSF procedure. The starting procedure CF corresponds to the generation of the initial state of positions over the target space (F), which can be randomized or guided by the connectivity information of the network (C). The repeating procedure starts from the current positions in the target space (F), which are then coupled with the connectivity information to calculate the forces among vertices. As a result a pairwise map of forces among vertices is obtained, which can be regarded as a S representation of the system. Finally, such information is used to calculate the new positions in the target space (F). The entire procedure is repeated until achieving a satisfactory embedding for the network in the target space (F).

\subsubsection{Spectral embedding}
\label{sec:spectralEmbedding}
The spectrum of a network can also be used to compute a reasonable embedding. Hall was one of the first to propose a technique for embedding a graph using the spectrum of the Laplacian matrix~\cite{hall1970r}. Recently, these techniques have been revisited and applied to visualize complex networks~\cite{koren2003spectral,koren2005drawing}.

The method proposed by Hall~\cite{hall1970r} is based on the spectral analysis of the quadratic form of the Laplacian matrix. The Laplacian matrix $\mathcal{L}$ of an unweighted network $\mathcal{G}(\mathcal{N},\mathcal{E})$ is defined as
\begin{equation}
\label{eq:LaplacianMatrix}
	\mathcal{L}_{ij} =
		\begin{cases}
			k_i ,	&	\text{if } i=j\\	
			-1, 	&	\text{if } i \neq j \text{ and } (i,j) \in \mathcal{E}\\
			0, 	&	\text{otherwise}
		\end{cases},
\end{equation}
where $k_i$ is the degree of node $i$. The quadratic form of the Laplacian matrix, $\mathbf{x}^T\mathcal{L}\mathbf{x}$ for a vector $\mathbf{x}=(x_0, x_1, x_2, ... x_N) \in \mathbb{R}^{|\mathcal{N}|}$ is given by
\begin{equation}
\label{eq:LaplacianQuadraticForm}
	\mathbf{x}^T\mathcal{L}\,\mathbf{x} = \sum\limits_{(i,j)\,\in\,\mathcal{E}}{(x_i-x_j)^2}.
\end{equation}

Hall suggested that a $1$-dimensional map could be generated from the minimization of Equation~\ref{eq:LaplacianQuadraticForm}, which bears similarities with the problem of finding a set of positions that optimize the distance among nodes. Additional constraints are nevertheless needed for this formulation. To avoid the trivial solution with all nodes lying in the same position, $\mathbf{x}$ must be orthogonal to $\mathbf{1} = (1,1,1...)$. The extension of the map to account for a target space of dimension $d$ is straightforward. This can be done by finding $d$ orthogonal vectors that minimize Equation~\ref{eq:LaplacianQuadraticForm} independently. It has been shown that the sequence of vectors that satisfies this condition is given by the eigenvectors of $\mathcal{L}$ corresponding to the lowest non-null eigenvalues. For instance, considering the case of a map to $\mathbb{R}^2$, and supposing that $\mathcal{L}$ has eigenvectors $(\mathbf{v}^{(0)},\mathbf{v}^{(1)},\mathbf{v}^{(2)},...\mathbf{v}^{(|\mathcal{N}|)})$ corresponding to the eigenvalues $0=\lambda_0<\lambda_1<\lambda_2<...<\lambda_{|\mathcal{N}|}$; the chosen positions for the nodes are given by the vectors $(\mathbf{v}^{(1)}$, $\mathbf{v}^{(2)})$. 

The computation of the eigenvectors of $\mathcal{L}$ is $O(|\mathcal{N}|^3)$ at the worst case~\cite{harel2002graph}. Nevertheless, many modifications of the visualization algorithm exist to account for smaller execution times, which, in general, avoid the procedure of calculating the entire set of eigenvectors of $\mathcal{L}$. Harel and Koren suggested some changes to the spectral method, defining a process called High-Dimensional Embedding (HDE)~\cite{harel2002graph}, which is based on the calculation of the spectrum only for some pivot nodes that are mapped to a space of much larger dimension than the target space. The resulting $m$-dimensional positions are projected via Principal Component Analysis (see Section~\ref{subsec:pca}) to the target space. The time complexity of the HDE method is $O(m^2|\mathcal{N}|+m|\mathcal{E}|)$. Considering these optimizations, the spectral techniques become much faster when compared to the force-directed approach. 

Spectral embedding can yield particularly good results for networks based on meshes and lattices~\cite{koren2003spectral,Spielman2007spectral}, however, since the minimizations of distances along the axis are taken independently, the final node positions may not reflect the distances among connected nodes, resulting in a layout of inferior quality when compared to force-directed methods. Nevertheless, a spectral layout can be used as the starting position for a force-directed method, so that the number of iterations of the algorithm to yield a good map is decreased~\cite{koren2002fast}. While the spectral method can be regarded as a CSF procedure, the topological distances among nodes are not directly involved in the calculation. As a consequence, it is more suitable to consider the spectral graph drawing as a direct CF procedure over the FCS space.

\section{Adjacency}

The adjacency transformation is the process of defining the connectivity of the network directly trough the features of the nodes, that is, without explicit considering their distances. Clearly, in some cases the distance will still be related to the presence of a connection, but the fundamental reason for the connection existence is not a short distance between the nodes. In the following we briefly present a few techniques aimed at providing the FC transformation. Some are based on the spatial position of the nodes, while others consider the \emph{connection attractiveness} of a node, which is related to the probability that the node will receive a connection.
\subsection{Tessellation networks}
\label{sec:Tesselation}

\begin{figure}[!htbp]
\begin{center}
\includegraphics[width=\linewidth]{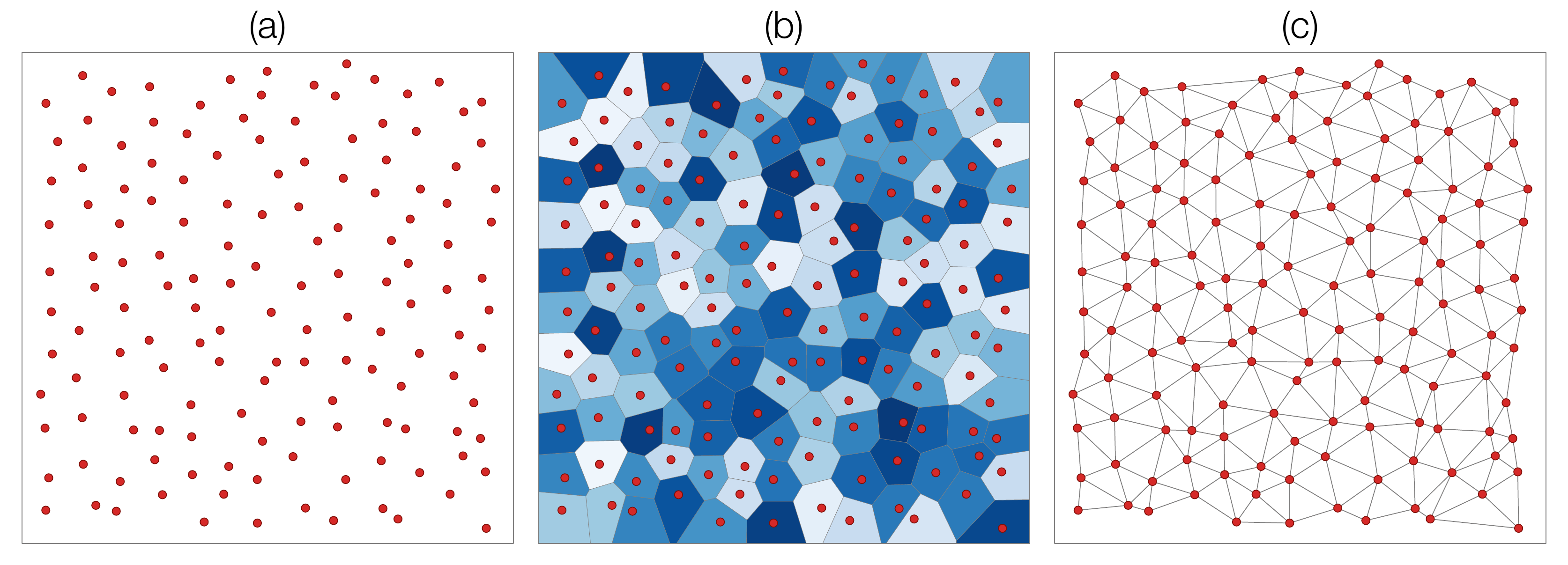}
\end{center}
\caption{Voronoi network construction. (a) Original set of points in a 2D space. (b) The respective Voronoi tessellation defined by the set of points. (c) The resulting network after connecting nodes belonging to adjacent Voronoi cells.}
\label{f:voronoi}
\end{figure}

Tessellation networks are defined from a process called \emph{spatial tessellation}~\cite{okabe2009spatial}. Nodes represent elements contained in a metric manifold and they are connected whenever their respective tessellation cells are adjacent. The most common kind of tessellation network corresponds to a Voronoi tessellation~\cite{aurenhammer2013voronoi} using the Euclidean distance as a metric. In Figure~\ref{f:voronoi} we illustrate how such networks are defined. From the original distribution of points in a 2D space (Figure~\ref{f:voronoi}(a)), a Voronoi tessellation is created, as shown in Figure~\ref{f:voronoi}(b). The set of positions that are closer to node $i$ than to any other node defines the Voronoi cell of node $i$. The network is then constructed by connecting nodes belonging to adjacent Voronoi cells, as seen in 
Figure~\ref{f:voronoi}(c). Such networks are equivalent to the Delaunay triangulation \cite{okabe2009spatial} of the set of points. Nevertheless, the Delaunay triangulation alone is not a suitable representation of such systems, since the cells associated with each node are usually also of interest for the analysis. For example, the size of the associated cell might be of importance to explain certain topological properties of the node. This means that the model can be applied as a FC transformation as well as a FCF transformation.

Tessellation models are useful in defining networks having strong spatial restrictions or cost on the connectivity, such as road, highway and power-grid networks. They also tend to provide a good representation of cellular structures such as biological tissue \cite{escudero2011epithelial,travenccolo2007new}. A related problem that bears similarities with tessellation models is granular packing \cite{arevalo2010topology,bassett2012influence,walker2010topological,walker2011complex,tordesillas2013revisiting}, in which a network of forces is used to describe interactions between adjacent particles. 

Scant attention has been given to tessellation networks. In part, this is a consequence of the great difficulty in developing analytical expression for properties of such networks. One of the few known analytical results is the probability $p_n$ for a Voronoi cell to have $n$ sides when the number of points is large and the points are placed according to a Poisson process \cite{hilhorst2005asymptotic,hilhorst2005perimeter}. This is given by \cite{hilhorst2005asymptotic}

\begin{equation}
p_n=\frac{C}{4\pi^2}\frac{\left(8\pi^2\right)^n}{\left(2n\right)!}\left[1+O\left(n^{-1/2}\right)\right],\label{eq:vorSides}
\end{equation}
where $C \approx 0.344$. Since the number of sides of a Voronoi cell is also the degree of the respective node, $p_n$ is also the degree distribution of the network. 

Another theoretical analysis done on Voronoi networks is the calculation of the site percolation threshold \cite{aharony2003introduction} of the model. In \cite{bollobas2006critical} it was found that the threshold is exactly $1/2$ for large networks. This means that when the probability $p$ of removing a vertex is $p<1/2$ a giant connected component exists in the network. Analytical results for the related problem of bond percolation are scarce. Nevertheless, in \cite{becker2009percolation} it was found that the bond percolation threshold is close, but not equal, to $1/3$. Note that the Voronoi network defined in \cite{becker2009percolation} is formed by the vertices of the Voronoi cells, but as shown in~\cite{bollobas2008percolation,becker2009percolation}, the bond percolation threshold of such network is dual to our definition of Voronoi network.

An interesting case where Voronoi tessellation was used as a tool for network analysis is presented in \cite{balcan2009multiscale}, where the authors created a Voronoi tessellation of the World Airport Network (WAN) \cite{guimera2005worldwide}. They used a gridded representation of the Earth surface, with a resolution of $15\times 15$ minutes of arc, and assigned each grid cell to the closest airport. They considered two restrictions for assigning cells. First, the grid cell had to belong to the same country of the airport. Second, the grid cell could not be farther than 200 km from the closest airport. Instead of considering the respective Delaunay triangulation of the airports, the authors used the calculated Voronoi cells to obtain the population density relative to each airport. This information was used to simulate epidemic dynamics \cite{barrat2008dynamical} in the associated airport network, where edges are related to direct flights between airports.
\subsection{Fitness networks}
\label{sec:Fitness}
The scale-free property is present in many real-world networks~\cite{redner1998popular,lawrence1999accessibility,albert1999internet,adamic2000power,liljeros2001web}. Initially, such characteristic seemed to be well explained in terms of the Barab\'asi-Albert (BA) model~\cite{barabasi1999mean,barabasi1999emergence}. The BA model is based on the idea that the potencial of a node acquiring new connections depends on the number of connections it already has. This is accomplished through two mechanisms: \emph{preferential attachment} and \emph{network growth}. However, the scale-free nature of many other real-world systems~\cite{adamic2000power,gonzalez_system_2006} can not be well explained by these two mechanisms alone. In some systems, the capacity of nodes acquiring new connections depends not only on the current topological state of the network but also on an intrinsic property of each node, known as \emph{fitness}. The fitness network model~\cite{bianconi2001competition} extends the BA model by establishing an interplay between fitness and preferential attachment, which gives rise to more realistic complex system models. In general, this happens because fitness represent a real-world information which is translated to the network topology and, as a consequence, defines the degree and other topological properties distributions. Therefore, the study of fitness networks can be understood as a FCF path.

The traditional procedure to generate a BA network~\cite{barabasi1999emergence} starts by constructing a small network comprising $m_0$ nodes. This network can be generated randomly (e.g, using the ER model) or by considering a complete graph. Next, $t$ new nodes are progressively added to the network. Each new node, is connected to $m$ (where $m \leq m_0$) nodes already present in the network, $G$, selected according to a probability proportional to the degree of each node $i$ according to 
\begin{equation}
\label{eq:fitness:newConnectionProbabilityBA}
	P_{k}^\text{\emph{new}} = {k_k \over \sum\limits_{j\,\in\,G} {k_j}}.
\end{equation}
This generates networks with older nodes presenting higher number of connections and overall degree distribution following a power law with exponent $\gamma = 3$ ($P_{\text{BA}}(k) \propto k^{-3}$). By eliminating any of the two main mechanisms of the model: preferential attachment or network growth, networks obtained with such procedure are no longer scale-free~\cite{barabasi1999emergence,albert2002statistical}. However, many real complex networks with heavy-tailed degree distribution cannot be fully described by the BA model. One limitation of the model is that it can only represent networks having exponent $\gamma = 3$, while real-world networks present a wide range of power law exponents~\cite{clauset2009power}. Also, there are networks where connectivity is not related to the age of a node, this is the case of the WWW, where no correlation between degree and age was found~\cite{adamic2000power}. Similarly, in sexual contact networks, the preferential attachment dynamics is not a satisfactory explanation for the emergence of scale-free distribution~\cite{gonzalez_system_2006}. In general, the number of previous sexual relationships of an individual is known a priori by the new partner, this mitigates the effects of any kind of preferential attachment based solely on the degree of nodes. A more reasonable explanation for the scale-free nature of sexual contacts is in the fact that the attractiveness of individuals is not evenly distributed. As a consequence, the distribution of the number of sexual contacts for individuals is also heterogeneous~\cite{gonzalez_system_2006,nguyen2012fitness}.

Bianconi and Barab\'asi introduced modifications to the standard BA model to account for some discrepancies with real networks~\cite{bianconi2001competition}. This was done by considering that nodes may present independent degree growth rates in addition to the preferential attachment mechanism. Thus, a fitness attribute is associated to each node and used to change its capacity to attract more connections. This new approach, known as \emph{fitness network model}, also presents the same mechanisms as the BA model, but instead of Equation~\ref{eq:fitness:newConnectionProbabilityBA}, the probability that node $i$ receives a connection depends on its fitness, $\eta_i$. In this context, Equation~\ref{eq:fitness:newConnectionProbabilityBA} is rewritten as
\begin{equation}
\label{eq:fitness:newConnectionProbabilityFitness}
    P_{i}^\text{\emph{fitness}} = {\eta_i k_i \over \sum\limits_{j\,\in\,G} {\eta_j k_j}}.
\end{equation}
It is interesting to analyze for which sets of fitness the resulting network yields a power-law degree distribution. This was studied by Bianconi and Barab\'asi\cite{bianconi2001competition} using continuous theory, where they also calculate the value of the respective power law exponent $\gamma$.

If the number of iterations, $t$, is large, the rate of growth of the number of connections of a node $i$ is given by the derivative of the continuous function $k_i(t)$. This derivative is proportional to the number of added edges per iteration, $m$, and the probability of connection at each iteration
\begin{equation}
\label{eq:fitness:degreeGrowthContinousFunction}
     {\partial k_i \over \partial t}(t) = m {\eta_i k_i \over \sum\limits_{j\,\in\,G} {\eta_j k_j}}.
\end{equation}
Equation \ref{eq:fitness:degreeGrowthContinousFunction} can be solved by considering that $k_i(t)$ follows a multi-scale power-law of exponent $\beta(\eta_i)$

\begin{equation}
\label{eq:fitness:degreeMultiscalePowerLaw}
     k_i (t) = m \left(t \over t_i \right)^{\beta(\eta_i)},
\end{equation}
with $t_i$ being the time at which node $i$ is added to the network.
From Equation~\ref{eq:fitness:degreeMultiscalePowerLaw} it is possible to show that $\beta(\eta)$ has the form~\cite{bianconi2001competition} 
\begin{equation}
\label{eq:fitness:betaExponentRelationship}
     \beta(\eta) = {\eta \over C},
\end{equation}
with $C$ being a normalization constant that can be obtained from the fitness distribution $\rho(\eta)$, using the normalization condition
\begin{equation}
\label{eq:fitness:fitnessNormalizationConstant}
     1 = \int_{0}^{\eta_{\text{max}}} \rho(\eta){\beta(\eta) \over {1-\beta(\eta)}} \, d\eta.
\end{equation}

The degree distribution of the model can be obtained from the cumulative probability of a node presenting $k_{\eta}(t) > k$
\begin{equation}
\label{eq:fitness:fitnessComulativeProbability}
	P(k_{\eta}(t) > k) = t\left({m \over k} \right)^{1 \over \beta(\eta)}.
\end{equation}
The probability density function for the vertices with fitness $\eta$ is then $P_\eta (k) = {d \over d k} P(k_{\eta}(t) > k) $. Thus, the overall degree distribution $P(k)$ can be written as a sum of all $P_\eta (k)$ weighted by the fitness distribution $\rho(\eta)$, that is
\begin{equation}
\label{eq:fitness:degreedistribution}
	P(k) = \int_{0}^{\eta_{\text{max}}} P_\eta (k) \,d\eta = \int_{0}^{\eta_{\text{max}}}  {\rho(\eta) \over \beta(\eta)} \left({m\over k}\right)^{{1 \over \beta(\eta)} +1} \,d\eta.
\end{equation}
When $\eta=1$ for all nodes, i.e. $\rho(\eta) = \delta(\eta-1)$, the problem reduces to the BA model with no fitness. In this case, $\beta=1/2$ which results in the known scaling law of the BA model, $P(k) = \alpha k^{-3}$. On the other hand, for a uniformly distributed set of fitness values, it can be shown~\cite{bianconi2001competition} that the degree distribution follows a different scaling law given by
\begin{equation}
\label{eq:fitness_degreedistributionUniform}
	P_\text{uniform}(k) \sim \alpha {k^{-\gamma^*} \over \log(k)}
\end{equation}
with $\gamma^* \approx 2.255$ . The logarithmic correction on Equation~\ref{eq:fitness_degreedistributionUniform} is related to the presence of vertices with very high number of connections (much higher than the number expected by the BA model). This effect is present on most real networks with competing elements such as the WWW and citation networks~\cite{bianconi2001competition,bianconi2001bose}.

In general, other fitness distributions can lead to largely distinct scaling laws and exponents. However, in \cite{bianconi2001bose} Bianconi et al. showed that well-behaved distributions of fitness can lead to three distinct network growth dynamics: ``rich-gets-richer'', ``fit-gets-richer'' or ``winner-takes-all''. The three scaling states can be understood in terms of a physical analogy to Bose gases. By assigning an energy value $\epsilon_i$ to each new vertex added to the network it is possible to redefine the fitness values of nodes in terms of their energy and a varying systemwide temperature parameter $T$
\begin{equation}
\label{eq:fitness_energy}
	\eta_j = e^{-{1\over T}\epsilon_j}.
\end{equation}
Thus, the fitness of a node decreases with its energy $\epsilon_j$ and is regulated by the term $1/T$.

When considering that $\epsilon_i$ follows a well-behaved distribution $g(\epsilon)$ and that $\lim_{\epsilon \to 0} g(\epsilon) = 0$, when $T\to \infty$, the fitness is the same for all vertices, resulting on the usual BA dynamics where older nodes are more likely to become hubs. Nevertheless, the temperature can be tuned in a way that the fitness become more important than the age of vertices. This happens when $T \ll 1$, yielding the ``fit-gets richer'' dynamics~\cite{ferretti2008dynamics}. In both cases, even with the presence of a few hubs, all vertices are only connected to a microscopic portion of the network in the continuum limit. However, it can be shown that for a critical temperature $T=T_c$ the system can attain a new state where a node connects to a macroscopic portion of the network on the continuum limit, resulting in a ``winner-takes-all" dynamics. This is remarkably similar to how non-interacting particles condensates at low temperature to form a Bose-Einstein condensate, thus, some researchers use the term ``Bose-Einstein condensate'' to specify this special case of network growth dynamics.

On spatial fitness networks~\cite{ferretti2011preferential}, the fitness values lie inside a hidden manifold. The model consists in uniformly populating a manifold with vertices and connecting them by preferential attachment, with fitness depending on the position of each node. The characteristic metric on the manifold generates various network behaviors such as the aforementioned growth dynamics. In particular, the aforementioned Bose-Einstein condensation occurs when the space contains singularities or high curvature (such as in hyperbolic spaces).

\subsection{Energy and fitness landscape networks}
\label{sec:EnergyAndFitnessLandscape}
Another important class of networks generated directly from the feature space is formed by the so-called \emph{energy landscape networks} (ELN)~\cite{stillinger1984packing,doye2002network,wales2003energy,massen2007preferential}. Such networks are generated from a potential energy function $\varphi(x_1, x_2, x_3, \dotsm x_n)$ lying in an Euclidean space or manifold of certain dimension $n$. Most ELNs present high adherence to its original space topology, thus they can be understood as geographic networks of very high dimension~\cite{doye2002network}.

\begin{figure*}[!tpb]
  \begin{center}
  \includegraphics[width=0.45\linewidth]{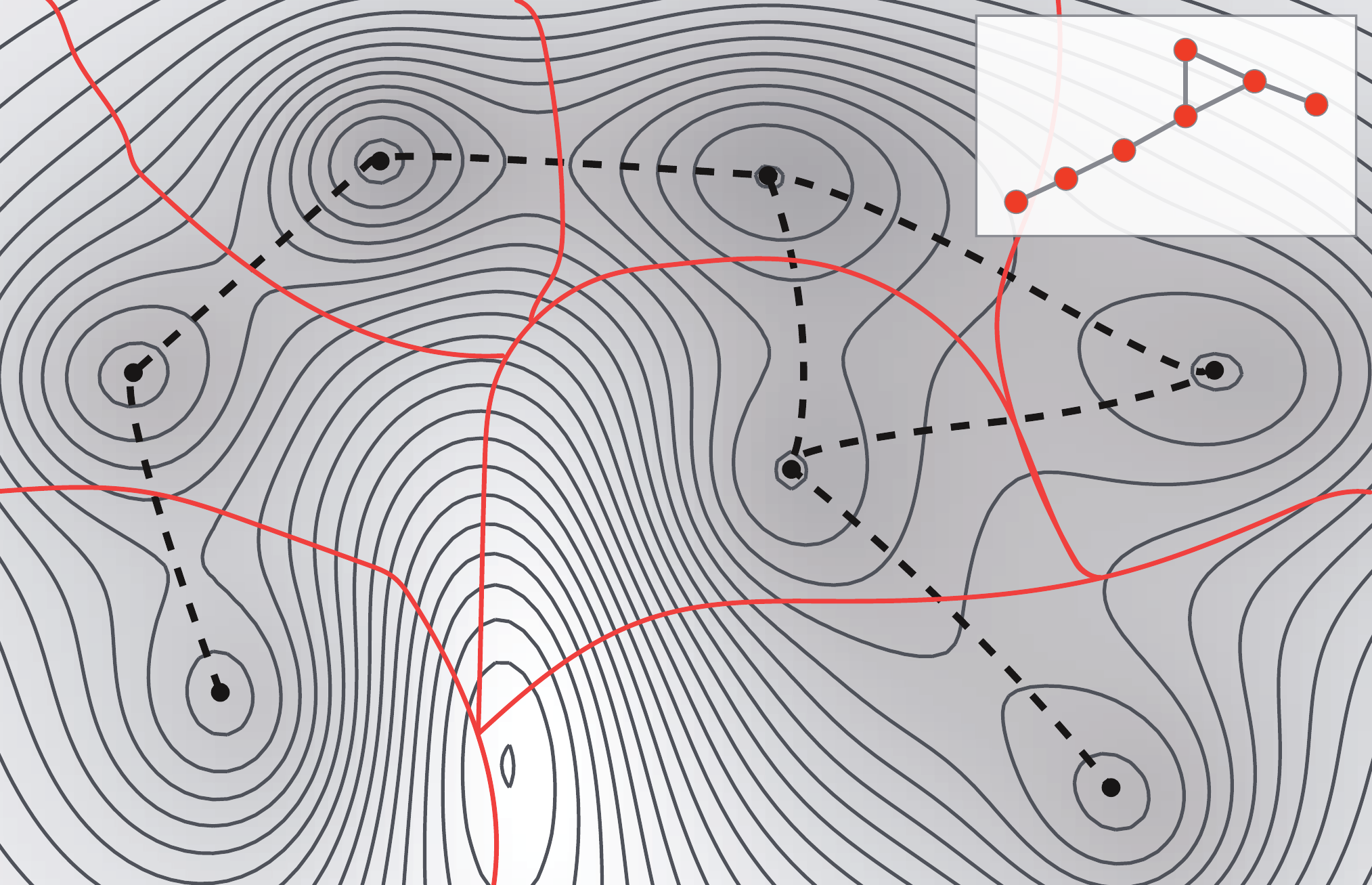} 
  \caption{Example of a ELN obtained from a 2D energy potential of two particles interacting by van der Waals forces. Equipotentials are displayed as full black lines and local minima of energy as black dots. The basins are shown in red while black dashed lines illustrate the possible phase transitions between states. The inset shows the network representation of this energy landscape.}
  ~\label{fig:potentialNetwork}
  \end{center}
\end{figure*}

ELNs are constructed considering each vertex as an attraction basin, $B=(b_1, b_2, \dotsm b_n$), corresponding to the regions around the local minima of a potential energy function $\varphi$. Two vertices $i$ and $j$ are connected whenever there exist a state transition from $i \to j$. Figure~\ref{fig:potentialNetwork} illustrates an ELN constructed from an energy landscape of interacting particles. A variety of methods can be used to obtain the basin structure of energy landscapes~\cite{tsai1993use,buchner1999potential,wales2003energy}. The methods differ according to the phenomenon and simulation techniques considered. Usually, a transition between two basins exists if there is a saddle point on the hyper surface between basins.

\begin{figure*}[!tpb]
  \begin{center}
\begin{center}
\subfigure[2D embedding]{\label{fig:generate_WS_1} \includegraphics[width=6.5cm]{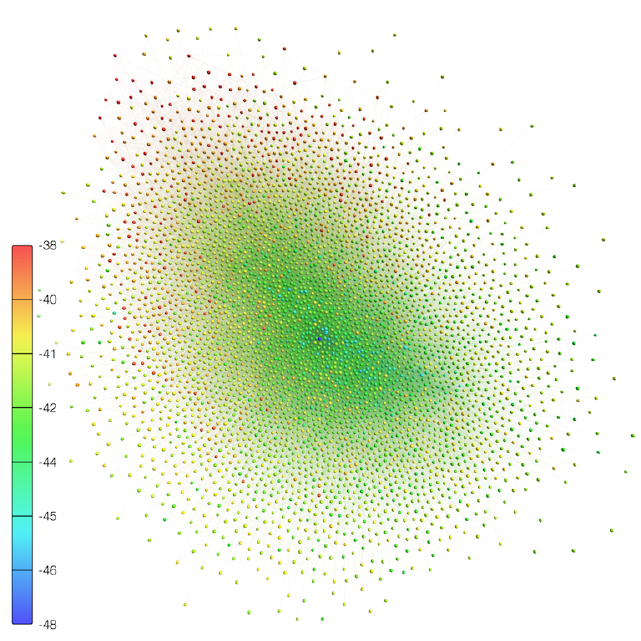} }~
\subfigure[Funnel structure]{\label{fig:generate_WS_1} \includegraphics[width=6.5cm]{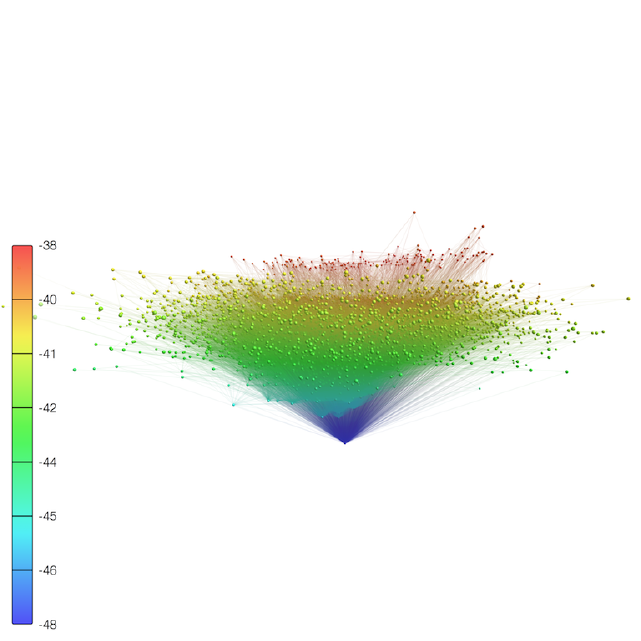} }
\end{center}
  \caption{ELN obtained from noble gases interacting by Leonard-Jones forces considering 14 particles. The network is shown as a 2D projection obtained by the FR algorithm (see Section~\ref{sec:forceDirectedtopologyEmbedding}) in {\bf a}, while {\bf b} highlights its characteristic funnel-like structure by using an extra dimension to account for the energy of the local minima. In both panels, the color maps the values of energy for each node. }
  ~\label{fig:funnelStructure}
  \end{center}
\end{figure*}

Doye and Massen observed that ELN constructed from simulated small Leonard-Jones clusters~\cite{tsai1993use} are scale-free. In addition, the degree of nodes in such networks correlates inversely with the energy of their respective basins~\cite{doye2002network}. As a result, low energy local minima tend to be highly reachable by presenting a wider range of available state transitions. This property is well-known in the energy landscape and protein fold fields as the \emph{folding funnel hypothesis}~\cite{onuchic1996protein,wales2003energy,oliveira2014visualization}. Figure~\ref{fig:funnelStructure} displays an ELN generated from simulated Leonard-Jones clusters and illustrates its funnel-like structure. 

Networks constructed from energy landscape are interesting because their topology depends solely on a steady potential energy of the state space. Therefore, although many ELNs exhibit scale-free degree distribution~\cite{doye2002network,rao2004protein,mousseau2005navigation,gfeller2007complex,hori2009folding}, neither growth dynamics nor preferential attachment mechanics are considered in their construction. This apparently contradicts the two prime aspects of how scale-free networks are constructed~\cite{massen2007self}. While the generation of scale-free networks with no growth dynamics are possible through a geographic network lying in a curved manifold (such as hyperbolic spaces, see Section~\ref{sec:spa_scale_free} for more details), this is not necessarily the case for ELNs. Its scale-free nature is better explained as a consequence of the fractal structure and hierarchical organization of its basins~\cite{massen2007exploring}. Apollonian networks~\cite{doye2005energy,doye2005self}, which are networks constructed by connecting adjacent elements in an Apollonian packing, have been proven to be good models for ELNs. The reason for the fractal organization exhibited in such systems are yet to be properly investigated. Similarly, ELNs derived from spin glasses systems also present fractal properties, which are related to the ruggedness of the energy surface~\cite{burda2007network,cao2015ground}.

The analysis of the community structure of ELNs revealed that such networks present very high modularity~\cite{massen2005identifying,cao2015ground}. This phenomenon is intuitively expected ELNs made of multiple funnels structures, with communities corresponding to the neighborhoods of each funnel. However, single funnel ELNs also yields very high modularity, but with no clear and natural indication of separation among the basins of the network~\cite{massen2005identifying}. Some recent developments indicate that, in general, networks lying in a manifold, such as geographical networks and the Apollonian graphs, tends to present high modularity~\cite{massen2005identifying,Arruda2016minimal,expert2011uncovering}.

The current investigations of ELN can be regarded as FCF class procedures. This is true for the analysis regarding the correspondence between topological features and characteristics of the basins, such as the tight relationship between node degree and the energy of local minima. The FCF class also holds for the visualization techniques employed to understand the funnel structure of ELNs. However, the investigations concerning the fractal nature of ELNs are regarded as a FCS procedure, since the information about the topological distance among basis becomes particularly important for such analysis.

\section{Conclusion}

Complex networks have widespread applications on many distinct areas, including pattern recognition, time series analysis and visualization. Nevertheless, many such applications are not readily known by researchers from these areas. In a similar fashion, many traditional tools developed for the analysis of real-world data are seldom used in network theory. The identification and integration of similar concepts studied in such distinct areas can greatly advance their development. Owing to this idea, our aim was to define three main representations of complex systems, and show that transformations applied to these representations can be associated with concepts usually found outside of network theory.

The \emph{feature}, \emph{similarity} and \emph{connectivity} representations of a system can be related through six immediate transformations, as shown in Figure~\ref{f:triangle}. A literature overview of each transformation was presented throughout the text, and some cases usually associated with network theory were illustrated. This allowed the organization of the six immediate transformations into a catalog of 42 possible transformation paths that can be applied to a system. The association of the presented methods with the considered paths allowed the identification of methodologies from different areas that have potential new applications. For example, the application of a closed path such as the FSCF can be used to identify the information loss of the network representation of a dataset. This could be done by comparing the original features with those obtained after projecting the network into a suitable space. Another interesting study revealed by the current framework is to combine distinct distance metrics with a number of edge selection techniques. We showed that the Euclidean distance followed by the exponential selection criterion defines the Waxman network model. It is an open question whether other traditional network models could be defined by additional metric-selection pairs.

\begin{table}[!htbp]
\tbl{Transformation paths considered throughout the text.}
{\begin{tabular}{ccc} \toprule
Path number	& Name & Count \\ \colrule
2  & FC   & 1 \\
4  & SF   & 1 \\
5  & CF   & 1 \\
7  & FSC  & 2 \\
8  & FCS  & 2 \\
12 & CSF  & 5 \\
13 & FSF  & 2 \\
14 & FCF  & 3 \\
18 & CSC  & 1 \\
19 & FSCF & 1 \\
\botrule
\end{tabular}}\label{tab:pathCount}
\end{table}

The identification of the 42 transformation paths also paves the way for new research about paths that have no associated methodology. In Table~\ref{tab:pathCount} we present the paths related to the methodologies discussed in the text. We also show how many methodologies were associated to each path. Clearly, it is very difficult to identify all methodologies related to each considered path. Still, it is clear that there remains many paths not studied in the literature. Such paths might reveal additional insights about complex systems that were not previously considered by researchers.

\section*{Acknowledgements}

C. H. Comin thanks FAPESP (grants no. 11/22639-8 and 15/18942-8) for financial support. T. K. D. M. Peron acknowledges FAPESP (grant no. 2012/22160-7) for support. F. N. Silva acknowledges CAPES and FAPESP (grant no. 15/08003-4). D. R. Amancio thanks FAPESP (grant no. 14/20830-0). F. A. R. acknowledges CNPq (grant no. 305940/2010-4) and 
FAPESP (grant no. 2013/26416-9). L. da F. Costa thanks CNPq (grant no. 307333/2013-2) and NAP-PRP-USP for support. This work has been supported also by FAPESP grant 11/50761-2.

\bibliographystyle{tADP}
\bibliography{references}

\end{document}